\documentclass[twocolumn]{aastex631}

\usepackage{wrapfig}
\usepackage{amsmath}
\usepackage{natbib}
\usepackage{graphicx}
\usepackage{appendix}
\usepackage{ulem}

\newcommand\msun{\hbox{\ensuremath{{\rm M}_{\odot}}}}

\newcommand{\scinot}[1]{\ensuremath{\times 10^{#1}}}
\newcommand{\mueff}{\ensuremath{\left<\mu\right>_{re}}}

\newcommand{\magas}{mag~arcsec$^{-2}$}

\begin{document}
\begin{abstract}
    A full understanding of how unusually large ``Ultra Diffuse Galaxies" (UDGs) fit into our conventional theory of galaxy formation remains elusive, despite the large number of objects identified locally. A natural extension of UDG research is the study of similar galaxies at higher redshift to establish how their properties may evolve over time. However, this has been a challenging task given how severely systematic effects and cosmological surface brightness dimming inhibit our ability to analyze low-surface-brightness galaxies at high-$z$. Here, we present a sample of low stellar surface density galaxies (LDGs) at moderate redshift, likely the progenitors of local UDGs, identified using deep near-IR observations of the El Gordo cluster at $z = 0.87$ with {\it JWST}. By stacking 8 NIRCAM filters, reach an apparent surface brightness sensitivity of 24.59~\magas{}, faint enough to be complete to the bright end of the LDG population. Our analysis identifies significant differences between this population and UDGs observed locally, such as their color and size distributions, which suggest that the UDG progenitors at high-$z$ are bluer and more extended than UDGs at $z=0$. This suggests that multiple mechanisms are responsible for UDG formation and that prolonged transformation of cluster dwarfs is not a primary UDG formation mechanism at high-$z$. Furthermore, we find a slight overabundance of LDGs in El Gordo, and, in contrast to findings in local clusters, our analysis does not show a deficit of LDGs in the center of El Gordo, implying that tidal destruction of LDGs is significant between $z=0.87$ and $z=0$.
\end{abstract}

\title[LDGs in El Gordo]{PEARLS: Low Stellar Density Galaxies in the El Gordo Cluster Observed with JWST}
\keywords{Low surface brightness galaxies, James Webb Space Telescope, Dwarf galaxies, Galaxy clusters, Galaxy structure, Stellar populations, Galaxy evolution}
\correspondingauthor{Timothy Carleton}
\email{tmcarlet@asu.edu}

\author[0000-0001-6650-2853]{Timothy Carleton} 
\affiliation{School of Earth and Space Exploration, Arizona State University,
Tempe, AZ 85287-1404, USA}

\author[0000-0003-3329-1337]{Seth H. Cohen} 
\affiliation{School of Earth and Space Exploration, Arizona State University,
Tempe, AZ 85287-1404, USA}

\author[0000-0003-1625-8009]{Brenda L. Frye} 
\affiliation{Steward Observatory, University of Arizona, 933 N Cherry Ave,
Tucson, AZ, 85721-0009, USA}

\author[0000-0001-9369-6921]{Alex Pigarelli} 
\affiliation{School of Earth and Space Exploration, Arizona State University,
Tempe, AZ 85287-1404, USA}

\author[0000-0002-3783-4629]{Jiashuo Zhang} 
\affiliation{Department of Physics, The University of Hong Kong, Pokfulam Road, Hong Kong}

\author[0000-0001-8156-6281]{Rogier A. Windhorst}
\affiliation{School of Earth and Space Exploration, Arizona State University,
Tempe, AZ 85287-1404, USA}

\author[0000-0001-9065-3926]{Jose M. Diego} 
\affiliation{Instituto de F\'isica de Cantabria (CSIC-UC). Avenida. Los Castros
s/n. 39005 Santander, Spain}

\author[0000-0003-1949-7638]{Christopher J. Conselice} 
\affiliation{Jodrell Bank Centre for Astrophysics, Alan Turing Building, 
University of Manchester, Oxford Road, Manchester M13 9PL, UK}

\author[0000-0003-0202-0534]{Cheng Cheng} 
\affiliation{Chinese Academy of Sciences, National Astronomical Observatories,
CAS, Beijing 100101, China}

\author[0000-0001-9491-7327]{Simon P. Driver} 
\affiliation{International Centre for Radio Astronomy Research (ICRAR) and the
International Space Centre (ISC), The University of Western Australia, M468,
35 Stirling Highway, Crawley, WA 6009, Australia}

\author[0000-0002-7460-8460]{Nicholas Foo}
\affiliation{Steward Observatory, University of Arizona, 933 N Cherry Ave,
Tucson, AZ, 85721-0009, USA}

\author[0000-0003-0883-2226]{Rachana A. Bhatawdekar} 
\affiliation{European Space Agency, ESA/ESTEC, Keplerlaan 1, 2201 AZ Noordwijk,
The Netherlands}

\author[0000-0001-9394-6732]{Patrick Kamieneski}
\affiliation{School of Earth and Space Exploration, Arizona State University,
Tempe, AZ 85287-1404, USA}

\author[0000-0003-1268-5230]{Rolf A. Jansen} 
\affiliation{School of Earth and Space Exploration, Arizona State University,
Tempe, AZ 85287-1404, USA}

\author[0000-0001-7592-7714]{Haojing Yan} 
\affiliation{Department of Physics and Astronomy, University of Missouri,
Columbia, MO 65211, USA}

\author[0000-0002-7265-7920]{Jake Summers} 
\affiliation{School of Earth and Space Exploration, Arizona State University,
Tempe, AZ 85287-1404, USA}

\author[0000-0003-0429-3579]{Aaron S. G. Robotham} 
\affiliation{International Centre for Radio Astronomy Research (ICRAR) and the
International Space Centre (ISC), The University of Western Australia, M468,
35 Stirling Highway, Crawley, WA 6009, Australia}

\author[0000-0001-9262-9997]{Christopher N. A. Willmer} 
\affiliation{Steward Observatory, University of Arizona, 933 N Cherry Ave,
Tucson, AZ, 85721-0009, USA}

\author[0000-0002-6610-2048]{Anton M. Koekemoer} 
\affiliation{Space Telescope Science Institute, 3700 San Martin Drive,
Baltimore, MD 21210, USA}

\author[0000-0001-9052-9837]{Scott Tompkins} 
\affiliation{School of Earth and Space Exploration, Arizona State University,
Tempe, AZ 85287-1404, USA}

\author[0000-0001-7410-7669]{Dan Coe} 
\affiliation{AURA for the European Space Agency (ESA), Space Telescope Science
Institute, 3700 San Martin Drive, Baltimore, MD 21210, USA}

\author[0000-0001-9440-8872]{Norman A. Grogin} 
\affiliation{Space Telescope Science Institute, 3700 San Martin Drive, 
Baltimore, MD 21210, USA}

\author[0000-0001-6434-7845]{Madeline A. Marshall} 
\affiliation{National Research Council of Canada, Herzberg Astronomy \&
Astrophysics Research Centre, 5071 West Saanich Road, Victoria, BC V9E 2E7, 
Canada}
\affiliation{ARC Centre of Excellence for All Sky Astrophysics in 3 Dimensions
(ASTRO 3D), Australia}

\author[0000-0001-6342-9662]{Mario Nonino} 
\affiliation{INAF-Osservatorio Astronomico di Trieste, Via Bazzoni 2, 34124
Trieste, Italy} 

\author[0000-0003-3382-5941]{Nor Pirzkal} 
\affiliation{Space Telescope Science Institute, 3700 San Martin Drive,
Baltimore, MD 21210, USA}

\author[0000-0003-0894-1588]{Russell E. Ryan, Jr.} 
\affiliation{Space Telescope Science Institute, 3700 San Martin Drive, 
Baltimore, MD 21210, USA}

\vspace{.5em}

\section{Introduction}
Low surface-brightness galaxies \citep{Impey1997} have intrigued astronomers for decades, with objects with $\mueff{}{}{}>24$~\magas{} having long been studied in nearby groups \citep[e.g.~][]{Jerjen2000}, clusters \citep[e.g.~][]{Impey1988,Thompson1993,Conselice2002,Conselice2003,Mieske2007,deRijcke2009,Penny2009,Penny2011} and the field \citep[e.g.~][]{Dalcanton1997}. 
Renewed interest in low-surface-brightness galaxies began with the identification of a large number of ``Ultra-Diffuse Galaxies" (UDGs) in Coma with the Dragonfly Telephoto Array \citep{vanDokkum2015}.
Since then, more UDGs have been identified in Coma \citep{Koda2015}, Virgo \citep{Mihos2015}, and Fornax \citep{Munoz2015}, as well as groups and the field \citep{Leisman2017,Roman2017a}. In addition to many objects identified in the local Universe, UDGs have also been identified in the Hubble Frontier Fields \citep{Janssens2017,Lee2017,Janssens2019,Lee2020,Ikeda2023} and $z\sim1$ clusters \citep{Bachmann2021}. Given their puzzling nature, optically-identified UDGs have been followed up with spectroscopy \citep{Gu2018,Ruiz-Lara2018,Ferre-Mateu2018,vanDokkum2019,Gannon2022}, X-ray \citep{Mirakhor2021}, UV \citep{Boissier2008}, IR \citep{Pandya2018,Buzzo2022}, and radio \citep{Leisman2017,Papastergis2017,Struble2018} observations across a wide range of environments \citep{Leisman2017,Prole2019}.

The large size of these galaxies (with half-light radii over $\sim1.5$ kpc) present challenges to theories of dwarf galaxy structure. Given the large globular cluster populations of many UDGs, some \citep[e.g.,][]{vanDokkum2016} suggest that UDGs are ``failed Milky Ways": galaxies with large halo masses and sizes that failed to form a correspondingly large stellar mass. Others suggest that UDGs live in typical dwarf halos that are expanded through star-formation feedback \citep{Chan2018} or high angular-momentum gas \citep{Amorisco2016}. Still others suggest that environmental effects such as ram-pressure stripping \citep{Junais2021,Safarzadeh2017}, galaxy harrasment \citep{Conselice2018} or tidal heating \citep{Ogiya2018, Bennet2018,Carleton2019} are responsible, given the large number of UDGs identified in galaxy clusters. The diversity of dynamical masses measured among  local UDGs, with some having unusually large dynamical masses \citep{vanDokkum2016,Beasley2016} and others having unusually small dynamical masses, even consistent with no dark matter \citep{vanDokkum2018,vanDokkum2019b}, makes distinguishing these theories with local observations challenging. Notably, these different theories make different predictions regarding the evolution of UDG abundance with redshift. If UDGs simply represent a fraction of normal dwarf galaxies, the ratio of UDGs to non-UDGs should be roughly constant out to $z>1$. On the other hand, if environmental processes are important for UDG formation, UDG formation may be less efficient at high redshift. In the local Universe, \cite{vanderBurg2017}, using clusters and groups with masses from $10^{12}-10^{15}~\msun{}$ identified in the GAMA survey \citep{Driver2011}, find a scaling relation of $N_{\rm UDG}=19\times\left(\frac{M_{200}}{10^{14}}\right)^{1.11}$, where $M_{200}$ is the mass within an overdensity where the average density is $200$ times the critical density. This suggests that the massive galaxy cluster ``El Gordo", or ACT-CL J0102-4915, identified as a Sunyaev–Zeldovich cluster at $z=0.87$ with the Atacama Cosmology Telescope telescope \citep{Menanteau2012}, with a halo mass of nearly $10^{15}~\msun{}$ should host approximately $226$ UDGs.

It is in this context that the El Gordo cluster was observed with JWST as part of the Prime Extragalactic Areas for Reionization and Lensing Science (PEARLS) GTO program (GTO 1176 \& 2738; PI Windhorst, Co-PI Windhorst \& Hammel; \citealt{Windhorst2023}). El Gordo contains two sub-clusters with masses of $4.6\pm0.77\scinot{14}~\msun{}$ and $4.7\pm0.77~\scinot{14}~\msun{}$ measured by applying the virial theorem to the radial velocities of the cluster members \citep{Frye2023}, and 1.38\scinot{15}~$h_{70}^{-1}$~\msun{} and 0.78\scinot{15}~$h_{70}^{-1}$~\msun{} from weak lensing analysis \citep{Jee2014}. This is somewhat in tension with the largest structures expected to form at this early time in $\Lambda$CDM. 
It was previously imaged in 10 filters with {\it HST} as part of the RELICS survey (GO 14096; \citealt{Coe2019}), which uncovered lensed arcs that supported measurements of its large mass \citep{Jee2014}. Given its young age (it is $\sim2$ Gyr younger than the highest-redshift Frontier Fields clusters), high mass (similar to Coma), and wealth of observations, El Gordo offers a prime target to study the evolution of cluster UDGs and UDG progenitors at high redshift.

In this paper, we characterize low surface density galaxies (LDGs) identified {\it JWST} images of the El Gordo Cluster. These objects include some of the first UDGs identified with JWST and are among the highest redshift cluster UDGs identified to date. In Section~\ref{sec:data}, we describe the observations of El Gordo taken with JWST and the process of data reduction and mosaic creation for those observations.
In Section~\ref{sec:identification}, we describe our procedure for identifying LDGs, as well as characterizing our completeness and our procedure for statistically subtracting contaminants. In Section~\ref{sec:results}, we describe the basic properties (size, stellar mass, and color) of these LDGs. In Section~\ref{sec:discussion}, we discuss our results in the context of UDG formation theories and in Section~\ref{sec:conclusion} we summarize our conclusions. Throughout, we state all sizes as circular half-light radii, and we assume a $\Lambda$CDM cosmology with $H_0=70$ km~s$^{-1}$~Mpc$^{-1}$, $\Omega_M=0.3$, and $\Omega_\Lambda=0.7$ \citep{PlanckCollaboration2020,Riess2022}. All magnitudes and surface brightnesses are given in the AB system \citep{Oke1983}.

\section{Data}
\label{sec:data}
\subsection{{\it JWST} Data}
{\it JWST} observed El Gordo as part of GTO program 1176 (PEARLS \citealt{Windhorst2023}) targeting moderate redshift clusters and blank fields to study galaxy evolution. El Gordo was imaged on July 29, 2022 with NIRCAM \citep{Rieke2023} SW filters F090W, F115W, F150W, and F200W in conjunction with LW filters F444W, F410M, F356W, and F277W (throughout, we refer to the NIRCAM module targeting the cluster center as the ``cluster field" and the additional module pointed 2.9\arcmin away from the cluster center as the ``parallel field"). Filters F090W, F115W, F444W, and F410M were observed with a MEDIUM8 readout with 6 groups per integration and 4 integrations for a total depth of 2491~s, filters F150W and F356W were observed with a SHALLOW4 readout with 9 groups per integration and 4 integrations for a total depth of 1890~s, and filters F200W and F277W were observed with a SHALLOW4 readout with 10 groups per integration and 4 integrations, for a total depth of 2104~s.

Images were first processed using the STScI {\sc calwebb} pipeline, with the {\sc jwst\_0995.pmap} context to perform basic photometric calibrations. After these initial calibrations, additional calibration steps were taken to remove artifacts in the image background. These corrections are described in detail in \cite{Windhorst2023}, but we briefly describe them here. First, ``1/f" noise (striping in the images due to amplifier differences) is corrected by matching the background level in adjacent columns using the {\sc ProFound} code \citep{Robotham2017,Robotham2018}. Similarly, detector-level offsets in the short-wavelength camera were corrected by constructing a ``super-sky" with {\sc ProFound} and bringing each detector onto that sky level. Lastly, ``wisps" caused by straylight from the secondary-mirror support were corrected using filter-specific templates.

In order to account for the large foreground population, we also utilize PEARLS observations of the North Ecliptic Pole Time Domain Field (TDF; GTO 2738), as the first epoch of observations was immediately released to the public to facilitate time-domain studies.
These deep observations provide an excellent dataset to characterize the foreground galaxy population and isolate the properties of LDGs in El Gordo. In particular, many dwarf galaxies in the El Gordo field are expected to lie in the foreground of the cluster, and given their closer distance, appear as LDG contaminants. By conducting our analysis on TDF data alongside the El Gordo cluster data, we can statistically subtract these contaminants and study the distribution of LDGs belonging to the cluster.

The raw data presented in this paper were obtained from the Mikulski Archive for Space Telescopes (MAST) at the Space Telescope Science Institute. The specific observations analyzed are archived here: \dataset[DOI: 10.17909/n1tn-bq90]{https://doi.org/10.17909/n1tn-bq90}. The observations of the TDF are publicly available as of this publication, and the observations of El Gordo will be publicly available beginning July 29, 2023.

\subsection{Local Cluster Data}
\begin{figure*}
    \centering
    \includegraphics[width=1\linewidth]{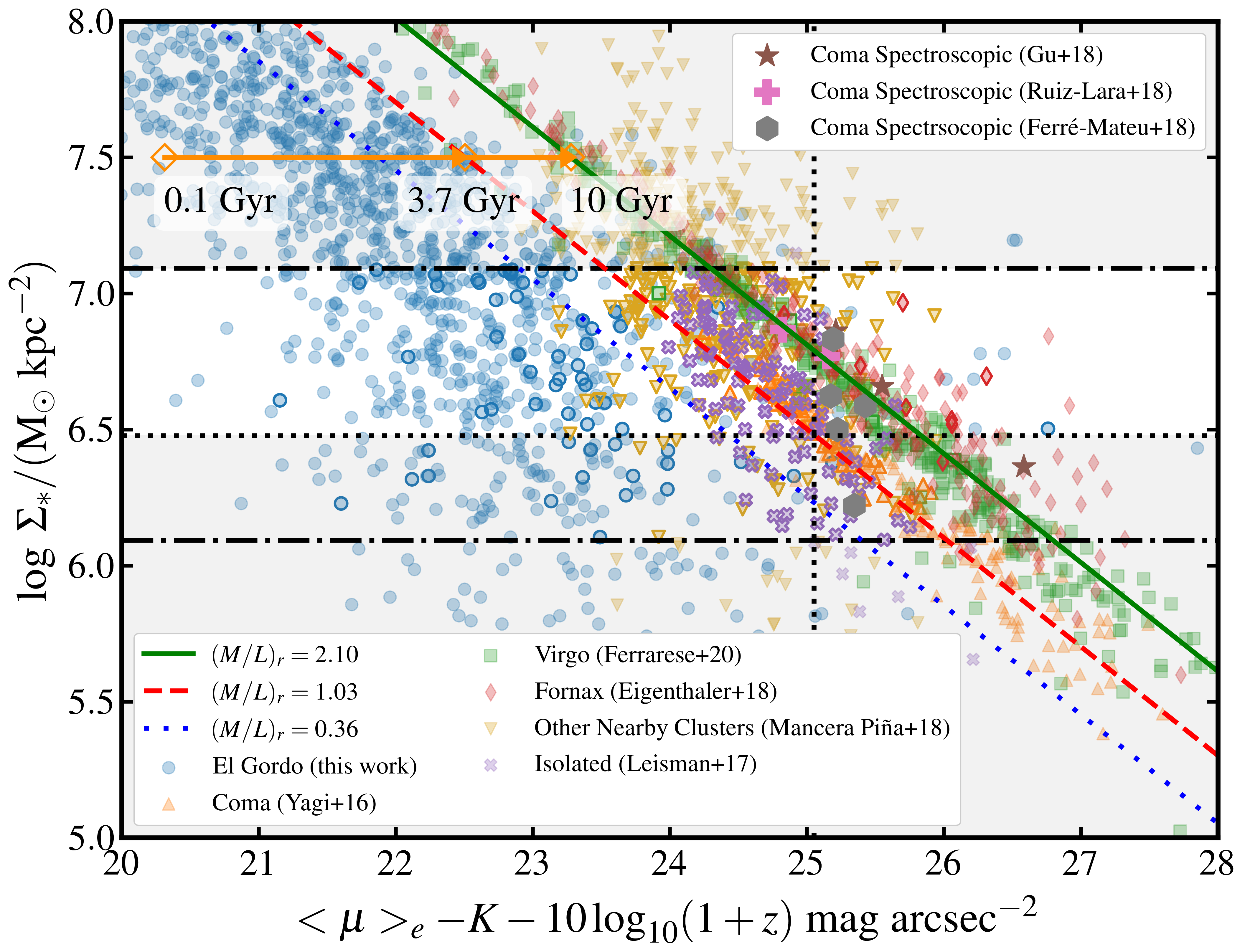}
    \caption{The relationship between rest-frame $r_{\rm CFHT}$-band surface brightness (\magas{} corrected for cosmological surface brightness dimming and $K$-corrected to $z=0$) and stellar mass surface density ($M_\odot$~kpc$^{-2}$) for samples of local UDGs and objects identified in El Gordo. Spectroscopic studies \citep{Gu2018,Ferre-Mateu2018,Ruiz-Lara2018} shown as large stars, hexagons, and ``+" signs, find that UDGs are largely composed of old, metal-poor stellar populations with a median mass-to-light ratio of $2.1$ (shown as the green solid line). Broader local cluster samples from Coma \citep[][orange upward triangles]{Yagi2016}, Virgo \citep[][green squares]{Ferrarese2020}, Fornax \citep[][red diamonds]{Eigenthaler2018}, other nearby clusters \citep[downward yellow triangles][]{ManceraPina2018} and the field \citep[][purple ``x"s]{Leisman2017} are also shown, with $B-V$, $g-r$, and $g-i$-based mass-to-light ratios derived from \cite{Bell2003}. Furthermore, our El Gordo measurements, which utilize SED-based stellar mass estimates, are shown as blue points, with the blue loosely dotted line corresponding to the average mass-to-light ratio of 0.58. Points with solid outlines are classified as LDGs. The black dash-dotted lines correspond to the stellar mass surface density limits described in Sec~\ref{sec:identification} and the dotted lines correspond to the completeness limits of our analysis. Notably, LDGs in El Gordo have significantly lower mass-to-light ratios than local objects. The orange line and points correspond to the evolution of a metal-poor stellar population from $0.1$ to $3.7$ to $10$ Gyr. The $10$~Gyr stellar population is consistent with the spectroscopic mass-to-light ratios, so the $3.7$~Gyr M/L ratio of 1.03 represents a reasonable upper limit of what we can expect for LDGs in El Gordo.}
    \label{fig:samplemasstolight}
\end{figure*}

To compare our results with clusters at $z=0$, we utilize catalogs published by the Next Generation Fornax Survey \citep{Munoz2015,Eigenthaler2018} and the Next Generation Virgo Survey \citep{Ferrarese2012,Ferrarese2020}. These surveys target the centers of the respective clusters and are sensitive to surface brightnesses of approximately $28$ and $27.2$~mag~arcsec$^{-2}$ respectively, such that they are fully able to identify the UDG population typically selected as having average $r$-band surface brightness between $24$ and $26.5$. Notably, these catalogs also contain higher surface brightness objects as well, to sample the full dwarf galaxy population. We also compare our results to the UDG catalogs from Coma \citep{Yagi2016}, nearby ($z<.04$) clusters \citep{ManceraPina2018,ManceraPina2019}, and the field \citep{Leisman2017}, although they do not contain objects with low stellar surface densities that have higher surface brightness (due to lower mass-to-light ratios).
Fig.~\ref{fig:samplemasstolight} shows these samples, as well as our objects, in $r$-band average surface brightness vs. stellar mass surface density space.

\section{LDG Identification}
\label{sec:identification}
Previous studies \citep[e.g.~][]{vanderBurg2016} have defined UDGs by both cuts in physical half-light radius ($>1.5$ kpc) and average surface brightness within $r_e$ ($24<\mueff{}{}{}<26.5$~\magas{}) based on ground-based optical imaging. As UDGs are identified in a wider range of redshifts with a wider range of instruments and detection bands, such cuts can result in an increasingly inhomogeneous sample \citep{VanNest2022}. For this reason, as well as to improve comparisons to models and simulations, we advocate studying galaxies in a range of \emph{stellar mass surface density} and physical size. {Note that this is largely consistent with a definition in terms of absolute surface brightness when assuming a particular mass-to-light ratio.} However, this is meant to represent a more physical quantity with a more standardized definition than surface brightness in an arbitrary band\footnote{Note that even among $r$-band surface brightness definitions, significant variation is present. For example, an old, metal-poor stellar population has a $\sim0.2$~mag difference between the Subaru $R_c$ band used in \cite{Yagi2016} and the CFHT $r$ band used in \cite{vanderBurg2016}.}. Specifically, we define LDGs as systems with stellar surface densities of $1.24\times10^6-10^7~{\rm M_\odot~kpc^{-2}}$. To convert the average surface brightness limits of $24-26.5$~\magas{} to circularized stellar mass surface density ($\Sigma_*=\frac{M_*}{2\pi r_e^2}$) limits, we assume a $R_c$-band mass-to-light ratio of $1.76$ (corresponding to an $r_{\rm CFHT}$-band mass-to-light ratio of $2.1$), taken as the median stellar mass-to-light ratio of spectroscopically characterized UDGs from \cite{Gu2018}, \cite{Ruiz-Lara2018} and \cite{Ferre-Mateu2018} with Subaru photometry from \cite{Yagi2016} (all corrected to Coma's luminosity distance of $102$~Mpc, and consistent with dwarf galaxies with old, metal-poor stellar populations). This is largely consistent with results from spectral energy distribution (SED) fits of local UDGs \citep[e.g.,~][]{Pandya2018,Buzzo2022,Webb2022}, although the possible impact of attenuation by dust on their stellar mass measurements means that we do not use them in our averaging process. We conduct SED fitting (Sec.~\ref{sec:meaesurement}) to determine the stellar masses of our objects, which are used to calculate the stellar surface densities for our selection. 
	
	While other studies \citep[e.g.~][]{Benavides2022, Li2023} define systems as outliers in the size-mass relation, we prefer a fixed stellar mass surface density and size criteria. We do so for 2 reasons: (1) given existing surface-brightness incompleteness in high-$z$ structural analysis, it is not clear what the size-mass relation at $z\sim1$ is for these dwarf galaxies, and (2) the physics of galaxy formation and evolution is more sensitive to the absolute surface density than relative size, so a surface-density based definition can be better compared to models and simulations.

\begin{figure}
	\centering
	\includegraphics[width=.9\linewidth]       {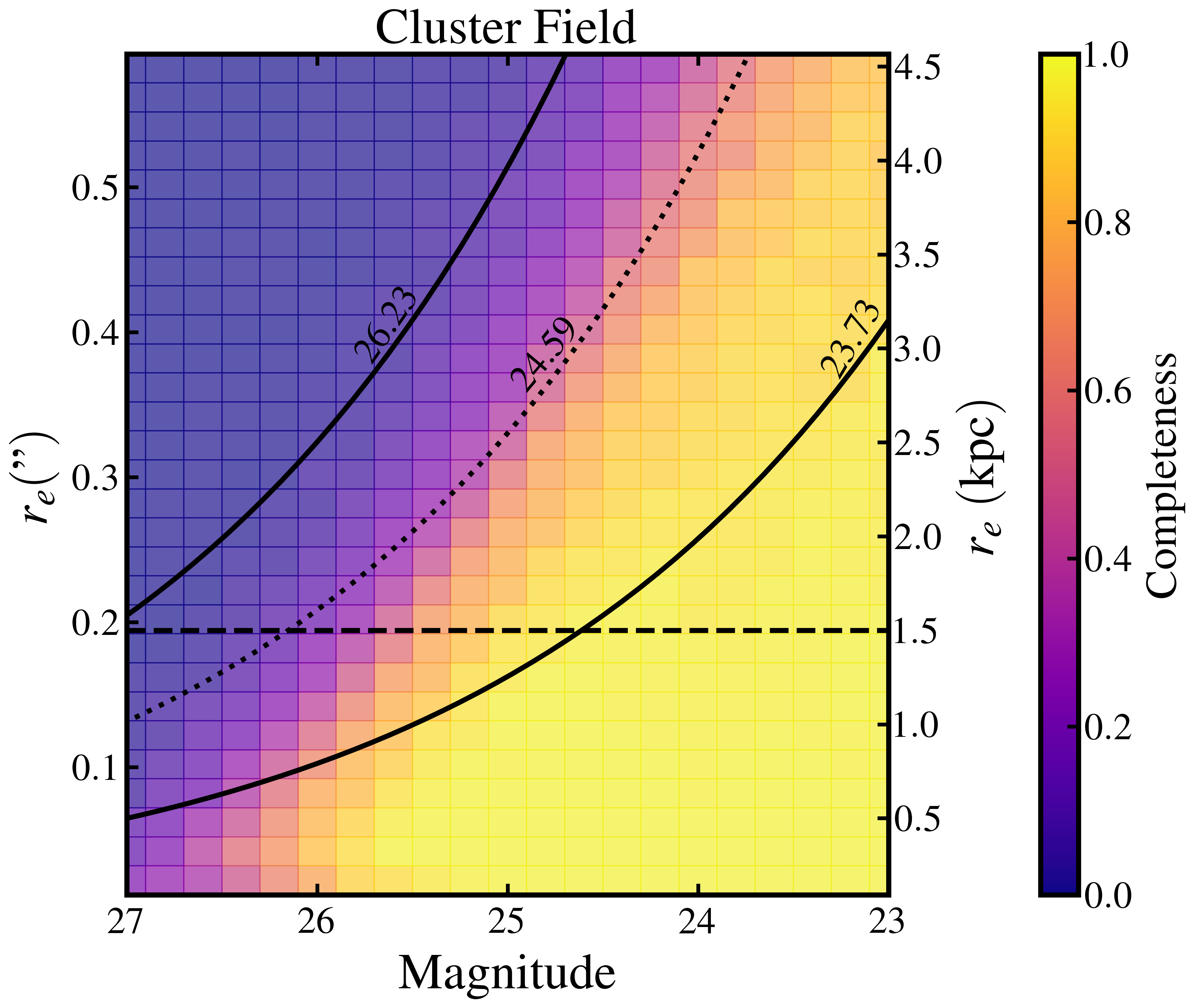}\\
        \includegraphics[width=.9\linewidth]{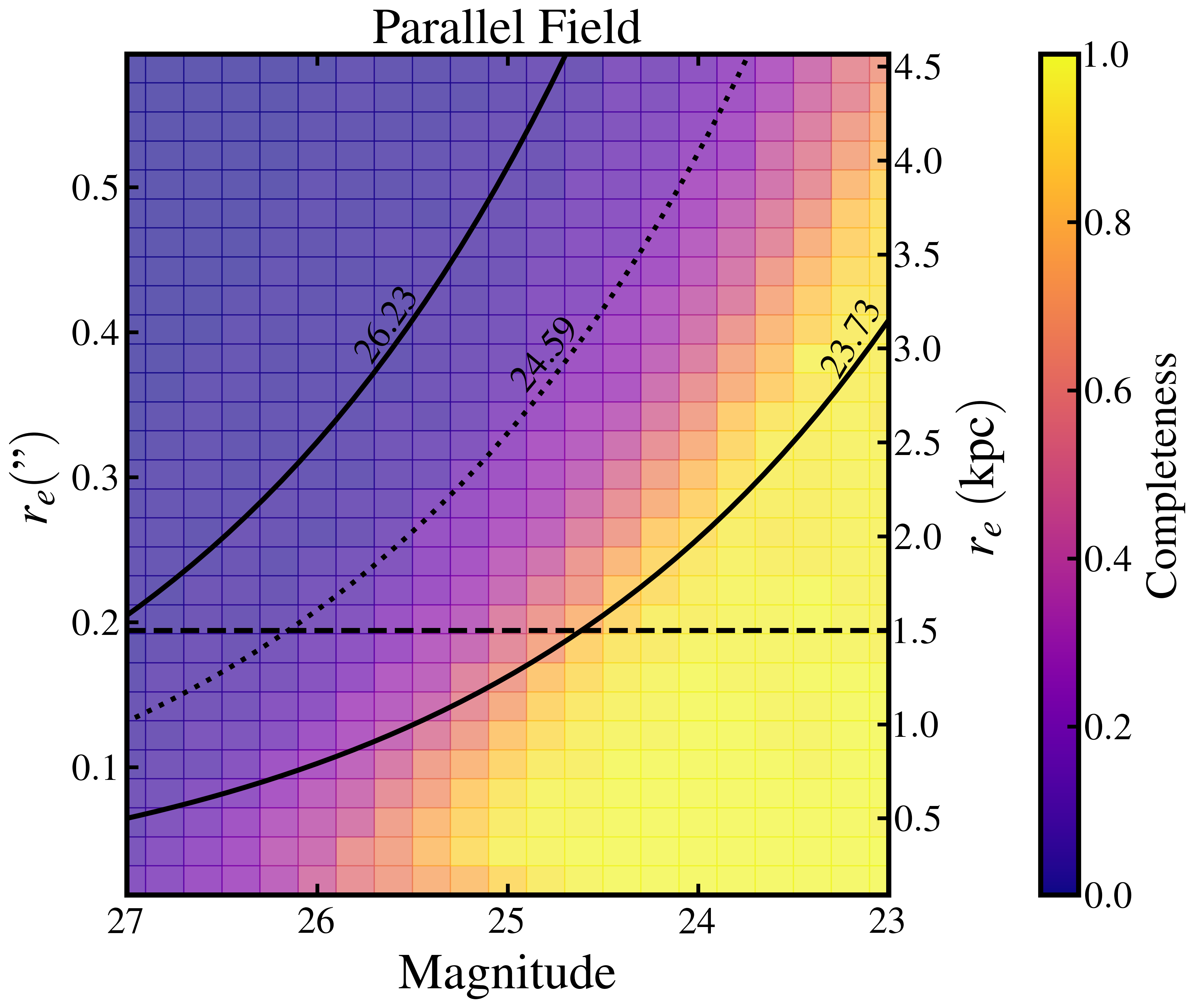}\\
        \includegraphics[width=.9\linewidth]{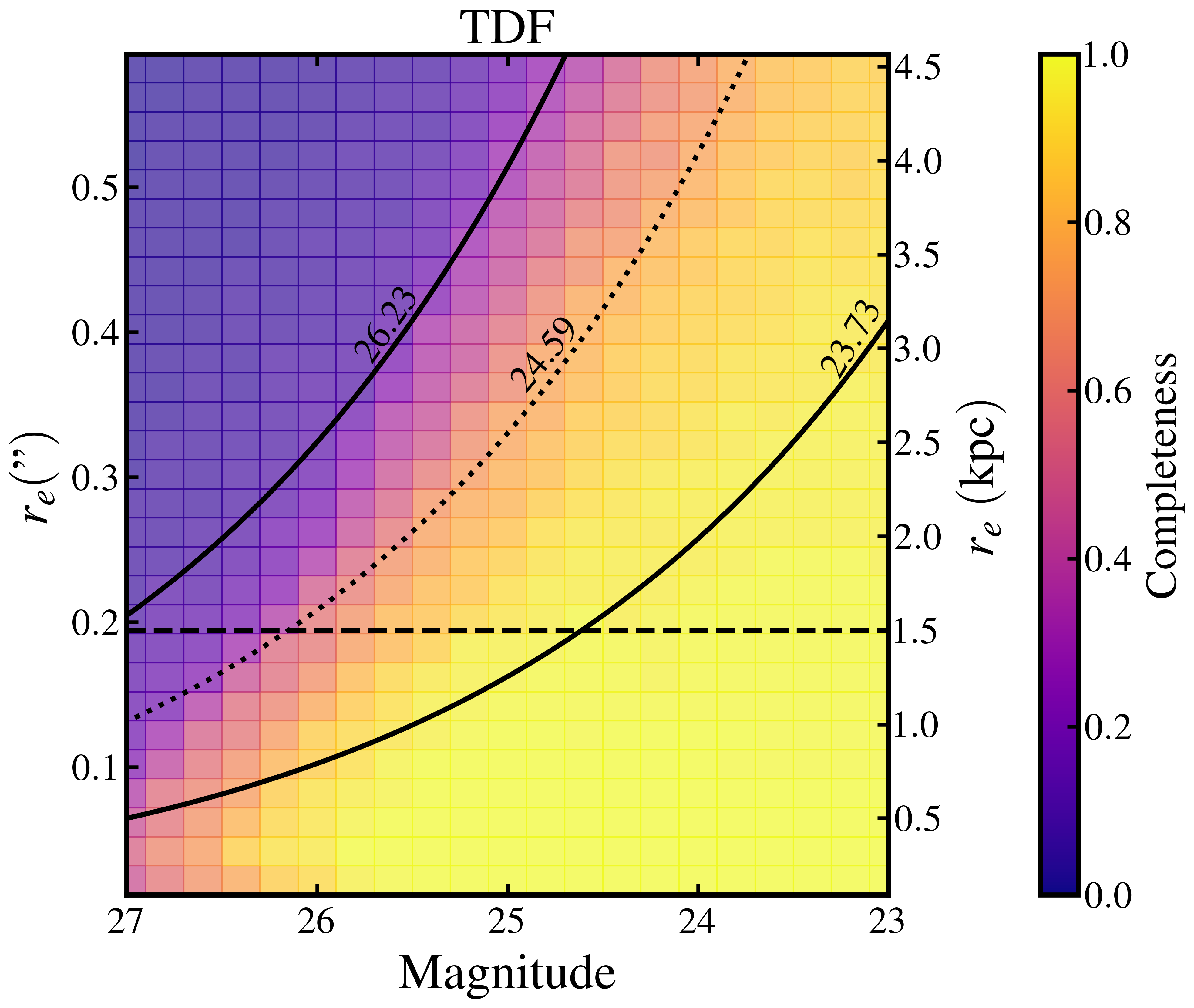}
	\caption{Our ability to detect objects in our stacked image as a function of total magnitude and half-light radius in the cluster field (top), parallel field (middle) and TDF (bottom). The horizontal dashed line at $0.19\arcsec{}$ corresponds to the physical size limit of $1.5$ kpc at $z=0.87$, and the solid curves correspond to average surface brightness limits of $23.05$ and $25.55$~\magas{}, the expected surface brightnesses of LDGs in our stacked image. Also, the dotted curve at $24.59$~\magas{} corresponds to the $50\%$ completeness limit in the stacked cluster image.}
	\label{fig:completeness}
\end{figure}

\subsection{Detection}
\label{sec:detection}
To identify LDGs, we stacked the images in all 8 NIRCAM filters, F090W through F444W (corresponding to rest-frame $g$ through $K$, drizzled to a pixel scale of $0.06\arcsec{}$). First, model PSFs were created with {\sc WebbPSF} v1.1.0, utilizing the {\sc odt} files closest to the observation date. Then, the images were convolved with a filter to match the F444W PSF. The matching kernel was constructed using the {\sc Photutils} PSF package \citep{Bradley2020}, which generates a matching kernel using a Fourier transform ratio, utilizing a split-cosine-bell windowing function to reduce high-frequency noise. Once the images were convolved, they were all stacked. We then run {\sc SourceExtractor} \citep{Bertin1996} on the PSF-matched, stacked mosaic to identify all sources in the image. We set {\sc BACK\_SIZE} to 1200, {\sc DETECT\_THRESH} to 0.9, and {\sc DETECT\_MINAREA} and {\sc ANALYSIS\_AREA} to 10 to optimize the detection of low-surface brightness objects. 

To test the detectability of LDGs in our imaging, we inject artificial sources, with a grid of magnitudes ranging from $22$ to $28$ and sizes ranging from $0.1\arcsec{}$ to $0.9\arcsec{}$, into our images and test our ability to recover inserted sources. The inserted galaxies were restricted to S{\'e}rsic indices of $1$ and axis ratios between $0.5$ and $1$ (consistent with observed UDGs; \citealt{vanDokkum2015}). The results of this analysis are shown in Figure~\ref{fig:completeness}.
Our apparent $50\%$ surface brightness limit in the cluster field (restricting to objects with half-light radii $>0.194\arcsec{}$) is $24.59$~\magas{}. To translate this to a surface density, we use {\sc python fsps} \citep{Conroy2009,Conroy2010,Foreman-Mackey2014} to model a single metal-poor ($\log Z/Z_\odot=-0.79$; taken as the median from the spectroscopic studies described above) stellar population with a formation redshift of $2.5$, consistent with the $\sim$ $10$~Gyr old stellar populations measured in local UDGs. At $z=0.87$, this corresponds to a $3.7$~Gyr old population with an $r_{\rm CFHT}$-band mass-to-light ratio of $1.03$ (higher than the 90th percentile of the SED-based mass-to-light ratios measured in our analysis below), which results in a surface density completeness limit of $0.30\scinot{7}$~{ \msun{}~kpc$^{-2}$. Throughout, we limit our analysis to objects above this surface density limit.

Also shown in Figure~\ref{fig:completeness} are black solid curves identifying \mueff{}{}{} of 23.05 and 25.55 ~\magas{}, which approximately correspond to the bright and faint ends of the LDG definition, and a dotted curve at 24.59~\magas{} corresponding to the $50\%$ completeness limit in the all-band cluster field. While we are not as sensitive as other studies at $z=0-0.5$ \citep[e.g.~][]{vanderBurg2017,Janssens2019}, we are able to sample the bright end of the LDG population at $z=0.87$.
Lastly, we note that the $2.7$ magnitudes of surface brightness dimming at the distance of El Gordo is apparent --- observationally, it required stacking 8 medium-deep {\it JWST} images to reach the depth achieved here. The use of near-IR observations helps because the mass-to-light ratio is about $5$ times higher in $i$ band (which roughly corresponds to rest-frame $g$-band at $z=0.87$) than F356W at this redshift.

\begin{figure}[h]
	\centering
        \includegraphics[width=1\linewidth]{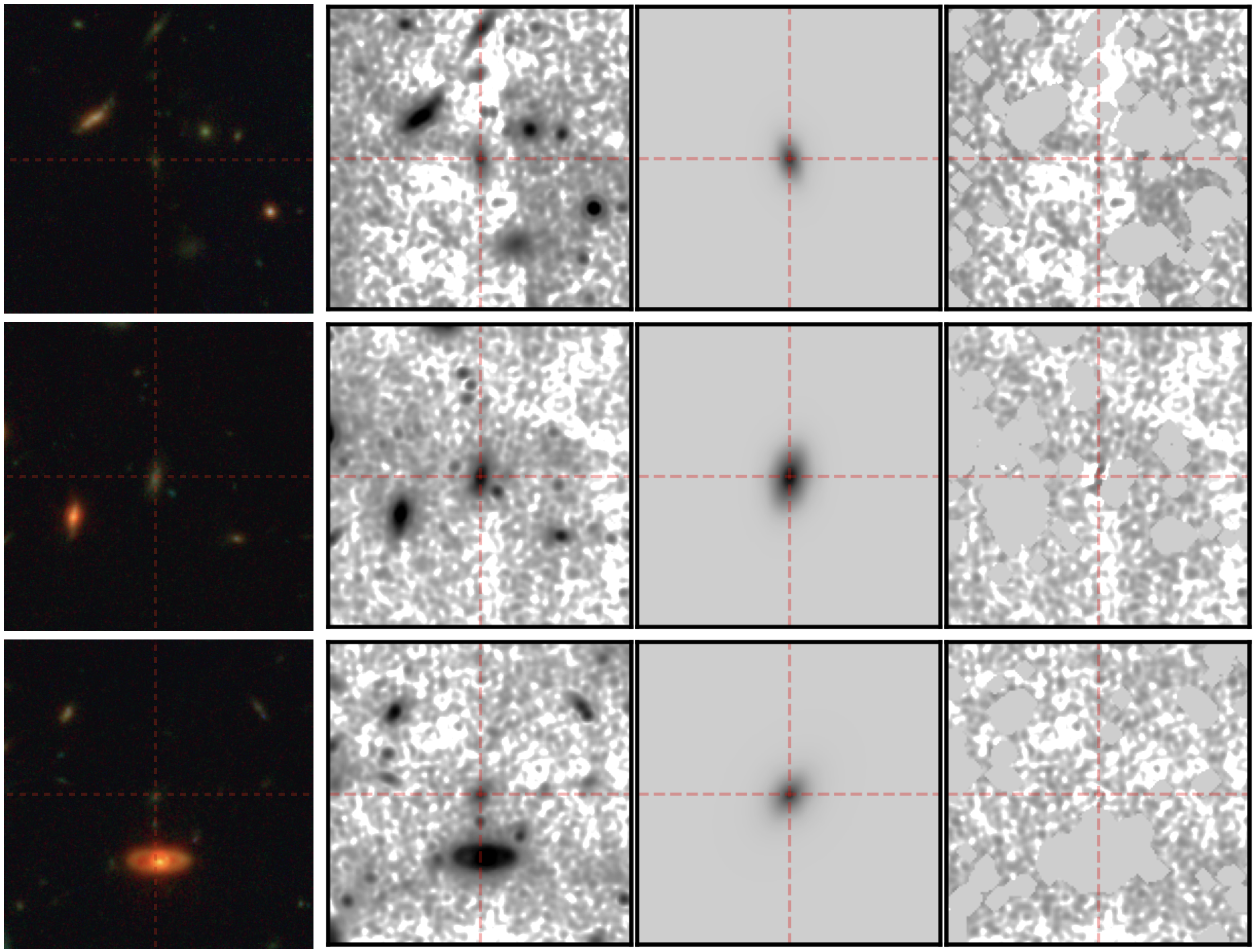}
	\caption{Example Galfit results on LDGs. {\bf Left:} A color image of the LDG (using F090W through F444W images). {\bf Middle Left:} The all-band cutout centered on the relevant object, with the background gradient subtracted (as described in Sec.~\ref{sec:identification}). {\bf Middle Right:} The best-fit {\sc Galfit} model. {\bf Right:} The residual image after subtracting the best-fit model, with other objects in the field masked.}
	\label{fig:example}
\end{figure}

\begin{figure*}
    \centering
    \includegraphics[width=1\linewidth]{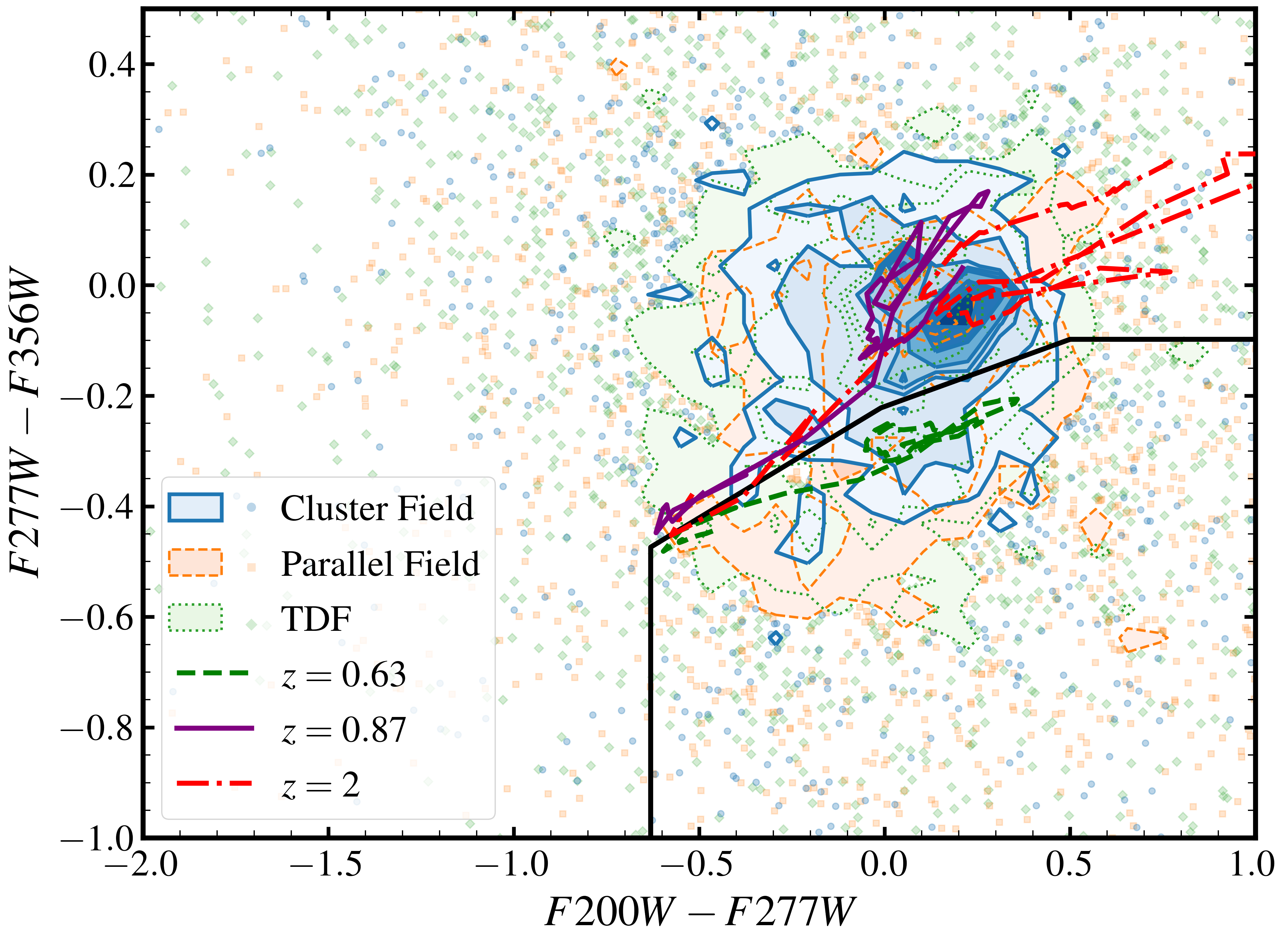}
    \caption{The $(F200W-F277W)$ and $(F277W-F356W)$ color-color space used to separate $z\sim0.63$ contaminants. Blue, orange, and green points are objects from the cluster field, parallel field, and TDF respectively. Green, purple, and red lines show galaxy tracks from {\sc FSPS} \citep{Conroy2009, Conroy2010} at $z=0.63$, $z=0.87$, and $z=2$, respectively, and the black line illustrates the color cut used to separate out $z\sim0.6$ contaminants. The extension of the cluster field contour near the $z=0.63$ track (many of which are otherwise classified as LDGs), separated from the rest of the objects, shows that this is effective at removing the low-$z$ contaminants. }
    \label{fig:colorcolor}
\end{figure*}

\subsection{Measurement}
\label{sec:meaesurement}
To identify LDGs, we measure the structural parameters of faint galaxies in our image with {\sc Galfit} \citep{Peng2010}. First, we identify dwarf galaxy candidates as  objects with magnitudes brighter than $30$~mag in the 8-filter stack (objects $>30$~mag are too faint to get reliable measurements), axis ratio greater than $0.2$ (to avoid spurious objects like diffraction spikes and the most obvious lensed galaxies), and away from any low exposure time edges of the image. Then, we create $250\times250$ pixel ($15\arcsec{}\times15\arcsec{}$) cutouts around those objects. To avoid contamination, we mask other sources within the cutout, growing the {\sc SourceExtractor} segmentation regions by $3$ pixels to avoid low surface brightness features contaminating the fit. Furthermore, we notice that the background varies on $\sim20-50$ pixel scales in our cutouts. After masking objects and dilating the masks, we used {\sc photutils MedianBackground} to map the background of the cutout on $20$ pixel scales. This background is then subtracted from the cutout. Although this effectively limits our ability to detect the largest UDGs ($10$~pix=$0.6\arcsec=4.6$~kpc), those objects are expected to be extremely rare (representing $<5\%$ of UDGs in local clusters). We then run {\sc Galfit} on all of the cutouts, using the {\sc SourceExtractor} MAG\_AUTO and {\sc FLUX\_RADIUS} as starting values for the total magnitude and half-light radius. Three example fits are shown in Figure~\ref{fig:example}. Given that UDGs and other dwarf galaxies have low S{\'e}rsic indices \citep{Koda2015, Roman2017a,Prole2019}, and background ellipticals are likely to have high S{\'e}rsic indices, we exclude objects with S{\'e}rsic indices greater than $2.5$. Additionally, we found that objects with sizes greater than $0.6\arcsec{}$ typically do not have reliable fits, so those objects are excluded from our sample. Given that these largest UDGs only represent $<5\%$ of UDGs in local clusters, we do not expect this selection to affect our conclusions. Lastly, we conduct a visual inspection of all fits to exclude the worst fits (e.g., obvious blending issues) from our sample. We are careful to only discard objects with severe issues, $\sim5\%$ of objects are discarded, in order to avoid biasing our selection process.

Notably, we assume single-component S{\'e}rsic profiles throughout this analysis. While many UDGs are observed to have bright nuclei \citep{Yagi2016} and some have faint tidal tails \citep{Bennet2018, Jones2021}, our assumption of single-component profile follows other studies of higher redshift UDGs \citep{Janssens2019, Lee2020} and is sufficient to identify LDGs based on average stellar density within $r_e$. Furthermore, UDGs are observed to have similar sizes in most bands \citep{Gannon2022}, such that our stacking procedure is not likely to affect our object identification.

\begin{figure*}[]
	\centering
	\includegraphics[width=1\linewidth]{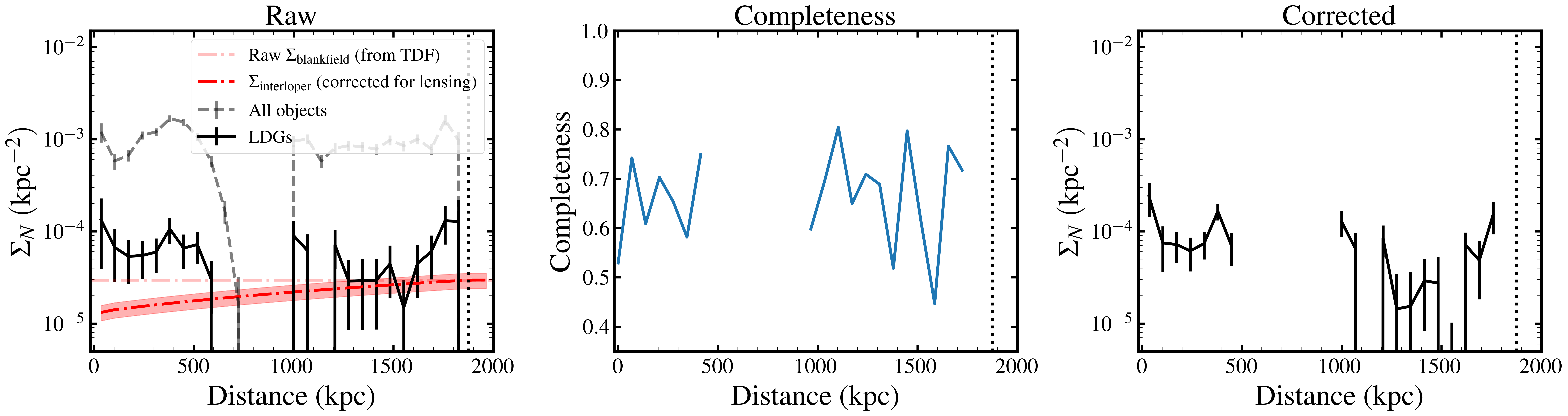}
	\caption{An illustration of the surface density of LDGs. {\bf Left:} Raw surface density of LDGs (solid) and non-LDGs (dashed) as a function of cluster-centric distance (taking the center to be J2000 R.A.: 01h02m55s, Dec: -49:15:33), as well as the foreground LDG surface density (taken from TDF data). The gap at 750-1100 kpc corresponds to the gap between the cluster and parallel fields. {\bf Middle:} Completeness as a function of cluster-centric distance. {\bf Right:} Surface density corrected for foreground contaminants and incompleteness.}
	\label{fig:densityvsr}
\end{figure*}
Once the structural parameters of all objects are measured, we estimate the stellar mass of our objects. First, we run {\sc SourceExtractor} in dual-image mode, with the stacked image as the detection image and the measurement image (convolved to the F444W PSF) set to the relevant filter to obtain multi-band photometry of our objects. Then we utilize {\sc Prospector} \citep{Johnson2021} to estimate the stellar mass based on the F090W through F444W photometry. While many objects are likely to be in the foreground or background, we enforce a redshift of $0.87$ because the signal-to-noise of the photometry is often not high enough to independently measure the redshift and stellar mass (the SED fitting doesn't have the benefit of stacking that allows us to measure structural parameters with {\sc Galfit}). For the SED fitting, we assume a delayed-tau SFH model, and we set uniform priors of log stellar mass $\in$ [6.5,10], age (Gyr) $\in{}$ [0,6.3] (corresponding to the age of the Universe at $z=0.87$), stellar metallicity ($\log Z/Z_{\odot}$) $\in{}$ [-1.5,0], and set E(B-V) to $0$ \citep[consistent with studies analyzing local UDGs;][]{Pandya2018}. While these fits will often give anomalous results for foreground or background objects (for example, the objects with extremely high mass-to-light ratios in Fig.~\ref{fig:samplemasstolight}), the procedure of analyzing the TDF in parallel allows us to statistically subtract these objects for the remainder of our analysis. In addition, the primary purpose of this SED fitting is to obtain stellar mass-to-light ratios for objects in our sample, so anomalous fits do not have a significant impact on our analysis.

After examining the colors of our sample, we imposed two additional color cuts to improve the purity of our sample. First, we apply a color cut of $(F200W-F356W)<0.25$ to exclude many high-redshift objects. This color represents the red edge of the red-sequence in the cluster field, and solar-metallicity objects are not expected to be redder than that color at $z=0.87$. In addition, a significant source of contamination is due to a $z=0.63$ group identified in the foreground of El Gordo by \cite{Caminha2022}. To best exclude these objects, we made cuts in $F(200W-F277W)$ and $(F277W-F356W)$ space (shown in Fig.~\ref{fig:colorcolor}), which is sensitive to the $1.6$ micron bump \citep{Sawicki2002} at $z=0.63$. A set of galaxies is present consistent with galaxies $z=0.63$, separate from the galaxies consistent with the cluster redshift of $z=0.87$. Objects with the following colors (consistent with stellar populations at $z=0.63$) are excluded from our sample:

\begin{align}
    F200W-F277W>&-0.631\label{eqn:colorcut}\\\nonumber
    F277W-F356W<&0.4058(F200W-F277W)-0.2179\\\nonumber
    F277W-F356W<&0.2431(F200W-F277W)-0.2196\\\nonumber
    F277W-F356W<&-0.098.\nonumber
\end{align}
We note that this color cut is indeed effective at reducing contaminants at with $z<0.75$. Only 4\% of objects with $z<0.75$ and $F115W<33$~mag in the CEERS mock catalogs (\citealt{Yung2022} and \citealt{Bagley2023}; based on the \citealt{Somerville2021} semi-analytic model) pass the color cut. In addition to removing the foreground group members from our sample and reducing the total number of contaminants, these cuts further alleviate the effect of cosmic variance on our study of LDG properties by narrowing the redshift range (and diversity of properties) of the contaminants.

We use these color cuts instead of photometric redshift cuts because the signal-to-noise of the measurements in the individual filters, as well as the limited wavelength range of our photometry (with only 1 filter blue-ward of the 4000\AA{} break), prevents us from being able to accurately measure photometric redshifts and stellar masses simultaneously. Furtheremore, our goal is to include as many cluster UDGs as possible --- the color cuts are more inclusive than uncertain photometric redshifts.

In summary, Table~\ref{tab:selectiontable} describes the criteria used to select objects in our parent sample. This is designed to be as inclusive as possible to real LDGs, while excluding obvious or significant contaminants. Of these, only objects that satisfy the stellar mass surface density and size cuts (summarized in Table~\ref{tab:udgselectiontable}) are classified as LDGs.

\begin{deluxetable}{|l|c|}
\tablenum{1}
\tablecaption{\label{tab:selectiontable} Galaxy Selection Parameters}
\tablehead{\colhead{Parameter} & \colhead{Range}}
\startdata
SE axis ratio & $>0.2$\\
SE MAG\_AUTO & $<30$ \\
$F200W-F356W$ & $<0.25$\\
    $F200W-F277W;F277W-F356W$ & Eqns.~\ref{eqn:colorcut}\\
    $r_e$ & $<0.6\arcsec$\\
    $n$ & $<2.5$\\
\enddata
\end{deluxetable}

\begin{deluxetable}{|l|c|}
\tablenum{2}
\tablecaption{\label{tab:udgselectiontable} LDG Selection Parameters}
\tablehead{\colhead{Parameter} & \colhead{Range}}
\startdata
  $\Sigma_*$ (${\rm M_\odot}~{\rm kpc^{-2}}$)  & $1.24\times10^6-1.24\times10^7$  \\
    $r_e$ & $>0.193\arcsec$\\
\enddata
\end{deluxetable}

\subsection{Contaminant Subtraction}
\label{sec:contaminants}
To estimate the contribution of foreground and background objects, we utilize PEARLS observation of the first spoke of the NEP TDF\footnote{We find that the upper half of our TDF image is not as complete as the lower half, likely because of difficulties in the sky subtraction. As a result,  we only use the lower half, corresponding to an area of 4.8 sq. arcmin. We utilize this field because it has the same filter sets and similar exposure times.}. We run the same procedure as described above, and identify $6.3$  galaxies per arcmin$^{2}$ that are within our selection criteria and above our completeness limit. The total LDGs abundance in the cluster is then calculated as:
\begin{equation}
    \begin{split}
        \Sigma_{\rm LDG} = \Sigma_{\rm LDG,~clust,~detect}(r)/C(r) -\\ \Sigma_{\rm LDG,~foreground}L(r)/C(b),
\end{split}
\end{equation}
where $C(r)$ is the completeness as a function of cluster-centric distance, $C(b)$ is the completeness in the TDF, and $\Sigma_{\rm LDG,~foreground}$ is the foreground LDG density, $\Sigma_{\rm LDG,~clust,~detect}$ is the spatial density of detected LDGs in the cluster field, and $L(r)$ accounts for lensing bias (see below). To calculate the spatially-variable completeness function, we insert 100 copies of each detected LDG from the TDF sample above our completeness limit throughout random positions within our stacked image and measure how many are recovered as a function of position. We find the spatial completeness increases from $0.5$ in the center to $0.7$ in the outskirts and parallel field. In all the distributions shown below, LDGs identified in the TDF are statistically subtracted from those in the cluster field to isolate the properties of the cluster LDGs.

Because of our complex selection criteria, the impact of cosmic variance on our results is not trivial. Regardless, as a first approximation, we use the ICRAR cosmology calculator\footnote{https://cosmocalc.icrar.org/} \citep{Driver2010} with a survey area of $4.8\arcmin$, a minimum redshift of $0.75$ (approximately the lower end of our Eqn.~\ref{eqn:colorcut} selection) and a maximum redshift of $4$ (approximately the upper end of our $F200W-F356W$ selection). The total cosmic variance error is $19\%$, which we show as a shaded region in Figure~\ref{fig:densityvsr}.

\subsubsection{Effect of Lensing}
\label{sec:lensing}
In addition to foreground contaminants, lensed background objects have the potential to be mis-identified as cluster LDGs. While our cuts in axis ratio and color exclude the most strongly lensed cases, we also model the expected abundance of lensed objects, to examine how they can affect our sample.  In particular, the true surface density of LDGs in the cluster ($\Sigma_{\rm cluster}$) is related to the observed surface density of LDGs ($\Sigma_{\rm observed}$) and the surface density of LDGs in a blank field ($\Sigma_{\rm blankfield}$) as:
\begin{align}
    \Sigma_{\rm cluster}=\Sigma_{\rm observed}/C(r) - \Sigma_{\rm blankfield}/C(b)\frac{B(r)+f}{1+f},
\end{align}
where $f$ is the ratio of foreground to background LDGs in the no-lens case and $B$ is the ratio of the abundance of background LDGs in the lens case to the abundance of background LDGs in the no lens case. To estimate $B$ and $f$, we utilize the stellar mass functions from COSMOS2015 \citep{Davidzon2017}, the late-type size-mass relation from \cite{vanderWel2014}, and the geometric lens model from \cite{Diego2023}. We restrict the redshift range to $z=0.75-4$ as a result of our color selections. We find $f=0.22$ and $B$ ranging from $0.22$ in the lensing center to $1$ in the outskirts, resulting in an average lensing correction factor of $0.52$. Alternatively, if we use a LTM model \citep{Frye2023}, the $B$ factors range from $0.3$ in the inner regions to $1$ in the outskirts. Notably, the strong lens results in substantial negative bias (consistent with magnification bias expected when sampling the faint end of the luminosity function, e.g.~\citealt{Umetsu2015}), but the relatively small uncertainty in the lens modeling does not significantly affect our results.

\section{Results}
\label{sec:results}
\begin{figure}
    \centering
    \includegraphics[width=1\linewidth]{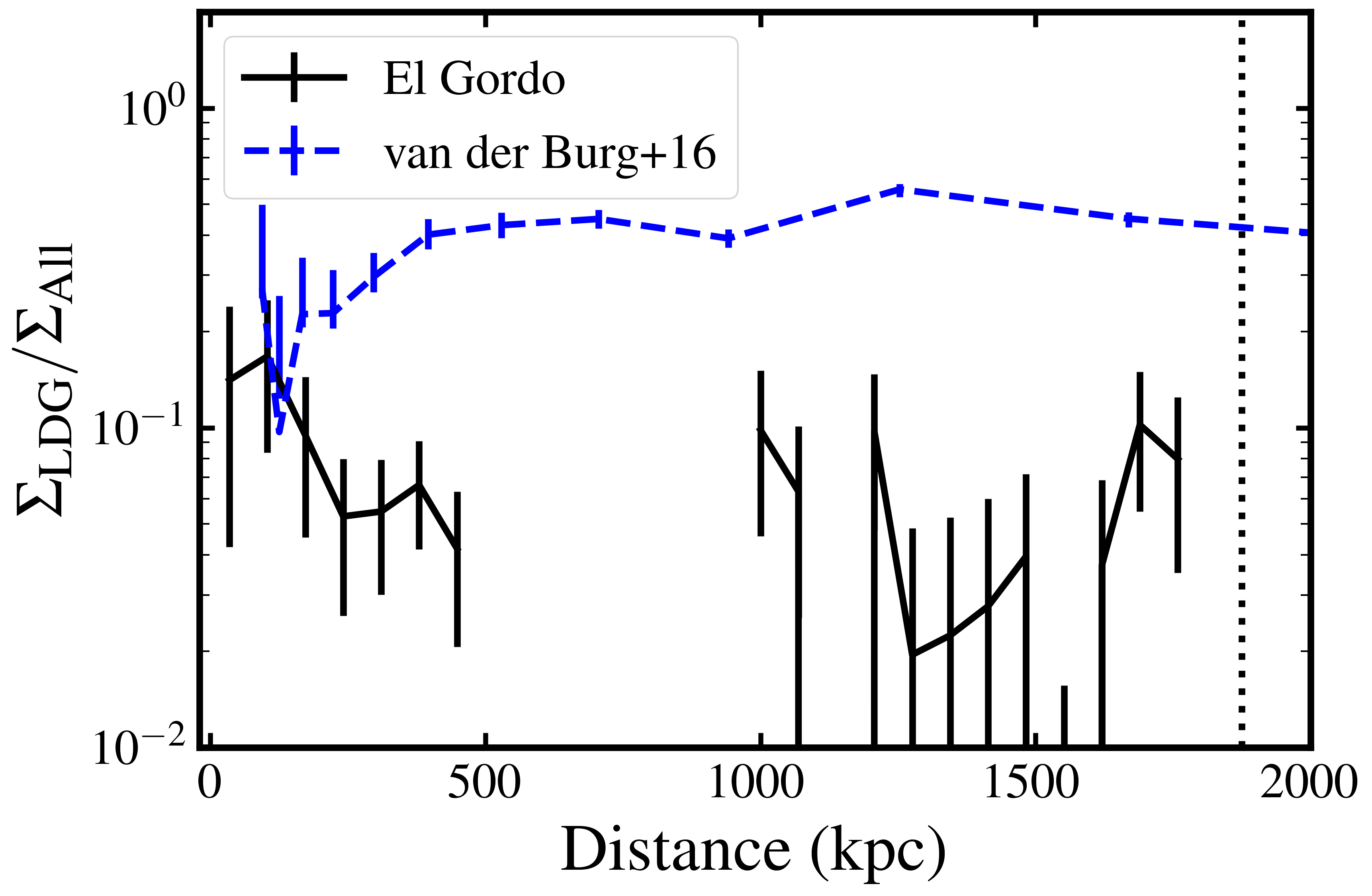}
    \caption{The ratio of the surface density of LDGs to total galaxies in El Gordo as a function of cluster-centric distance. As opposed to the ratio of UDGs to non-UDGs in local clusters \citep[e.g.~][~shown as the blue line]{vanderBurg2016}, we do not see a deficit of LDGs in the cluster center. This suggests that between $z=0.87$ and $z=0$, tidal effects have significantly disrupted the LDG population.}
    \label{fig:ratiovsr}
\end{figure}
\subsection{Spatial Distribution}
\label{sec:spatialdistribution}
Figure~\ref{fig:densityvsr} shows the surface density of LDGs as a function of distance from the El Gordo center (taken to be the light-weighted center of of the F444W image: 01h02m55s,-49:15:33). The first panel shows the raw surface density measurement, as well as the surface density of foreground objects (both raw from the TDF and corrected for lensing bias), the middle panel shows the completeness as a function of cluster-centric distance, and the right panel shows the LDG surface density corrected for incompleteness and foreground contamination. The surface density only decreases slightly from the center to the outskirts, largely reflecting the overall trend observed in dwarf galaxies in El Gordo. This LDG/non-LDG correspondence is reflected in Fig~\ref{fig:ratiovsr}, which shows the ratio of LDG to total galaxy surface densities as a function of cluster-centric distance. As seen in other studies \citep{vanderBurg2016}, we see LDG abundance largely mirroring the total galaxy abundance (which is shown as the dashed line in Fig~\ref{fig:densityvsr}a). 

However, a notable feature of UDGs in local clusters is a pronounced deficit in the cluster center, typically a factor of $\sim3$ (with the results from \citealt{vanderBurg2016} shown as the blue dashed line in Fig.~\ref{fig:ratiovsr}). This has commonly been explained as due to tidal effects destroying the LDG population more efficiently in the cluster center. We do not find such a deficit in El Gordo.\footnote{Note that if we center on the BCG in the southeast of the cluster, there may be evidence of a deficit of LDGs within $70$~kpc; however the uncertain incompleteness correction that close to the BCG makes this measurement particularly uncertain and in any case the deficit does not extend further.} Fig.~\ref{fig:ratiovsr} shows that UDGs are found at similar rates as non-UDGs throughout the cluster. Notably, \cite{Janssens2019}, studying UDGs in the Frontier Fields clusters, find a similar trend with redshift. The result of the study in the Frontier Fields clusters is that clusters at $z\sim0.3$ have a pronounced deficit of UDGs in their centers, whereas clusters closer to $z=0.5$ lack such a deficit. Our analysis, combined with their measurements, suggests that most of the tidal disruption of LDGs takes place between $z=1$ and $z=0$.

\subsection{Abundance}
\label{sec:abundance}
Following our selection procedure, we find $43$ LDGs in the El Gordo cluster field, and 33 LDGs in the parallel field above our stellar density completeness limit. Integrating the LDG surface density found in Sec.~\ref{sec:spatialdistribution} out to $1500$ kpc (where the abundance blends in with the background abundance), we calculate a total inferred LDG abundance of $439 \pm 112$ (this error includes the $19\%$ cosmic variance error identified in Sec.~\ref{sec:contaminants}). Assuming that our sample represents $64\%$ of the LDG population (the amount expected given the surface density distribution from the NGFS and NGVS), this corresponds to a total abundance of $686 \pm 175$. This is shown within the context of other UDG abundance measurements in Fig.~\ref{fig:nvsm}.
This is approximately a factor of $3$ higher than suggested by the scaling relation of \citep{vanderBurg2017}. In the limit that no more LDGs are accreted onto El Gordo, this suggests that fewer $40\%$ of the LDGs observed in El Gordo will survive until $z=0$, with most of the observed LDGs becoming part of the ICL.

As for the abundance of surface-brightness-selected UDGs in El Gordo, our results are more uncertain. If we select objects based on $r$-band absolute surface brightness ($<\mu>_{\rm abs}=<\mu>_e-10\log_{10}(1+z)-K-E$; \citealt{Graham2005}), where the $K$ correction from $F115W$ to $r$ is $K=0.143(F115W-F150W)-0.002-2.5\log(1.87)$ \citep{Hogg2002} and the evolutionary $K$-correction $E$ is set to $-0.90$~mag based on a $\log(Z/Z_{\odot})=-0.79$ $z_{\rm form}=2.5$ single stellar population), we find 25 UDGs in the cluster field. Extrapolating to the whole cluster area, and correcting for the foreground density of 2.5 per square arcmin, results in 283 UDGs. However, scaling this to the whole UDG population is more challenging. Whether UDG formation is taken to be at $z=1.5$, $z=2.5$, or $z=4$ results in evolutionary K-corrections of $1.2$, $0.9$, or $0.7$ respectively, and whether the stack-to-$r$ band K-correction is $-3.16$ as suggested by a single $z_{\rm form}=2.5$ stellar population or $-3.9$ (the median of our sample), means that our completeness limit of $24.59$ can correspond to an $r$-band absolute surface brightness limits ranging from $25.73$ to $26.97$, resulting in a completeness of $74\%$ or $100\%$. This suggests an overall UDG abundance ranging from $283-384$ (i.e. $283/1.00$ to $283/0.74$), close to the \citep{vanderBurg2017} scaling relation. Whether this is inconsistent with an of evolution in UDG abundance with redshift (as suggested by \citealt{Bachmann2021}) is not clear given the large scatter in UDG abundance at $z=0$ and the large uncertainties in our measurement.

\begin{figure}
	\includegraphics[width=1\linewidth]{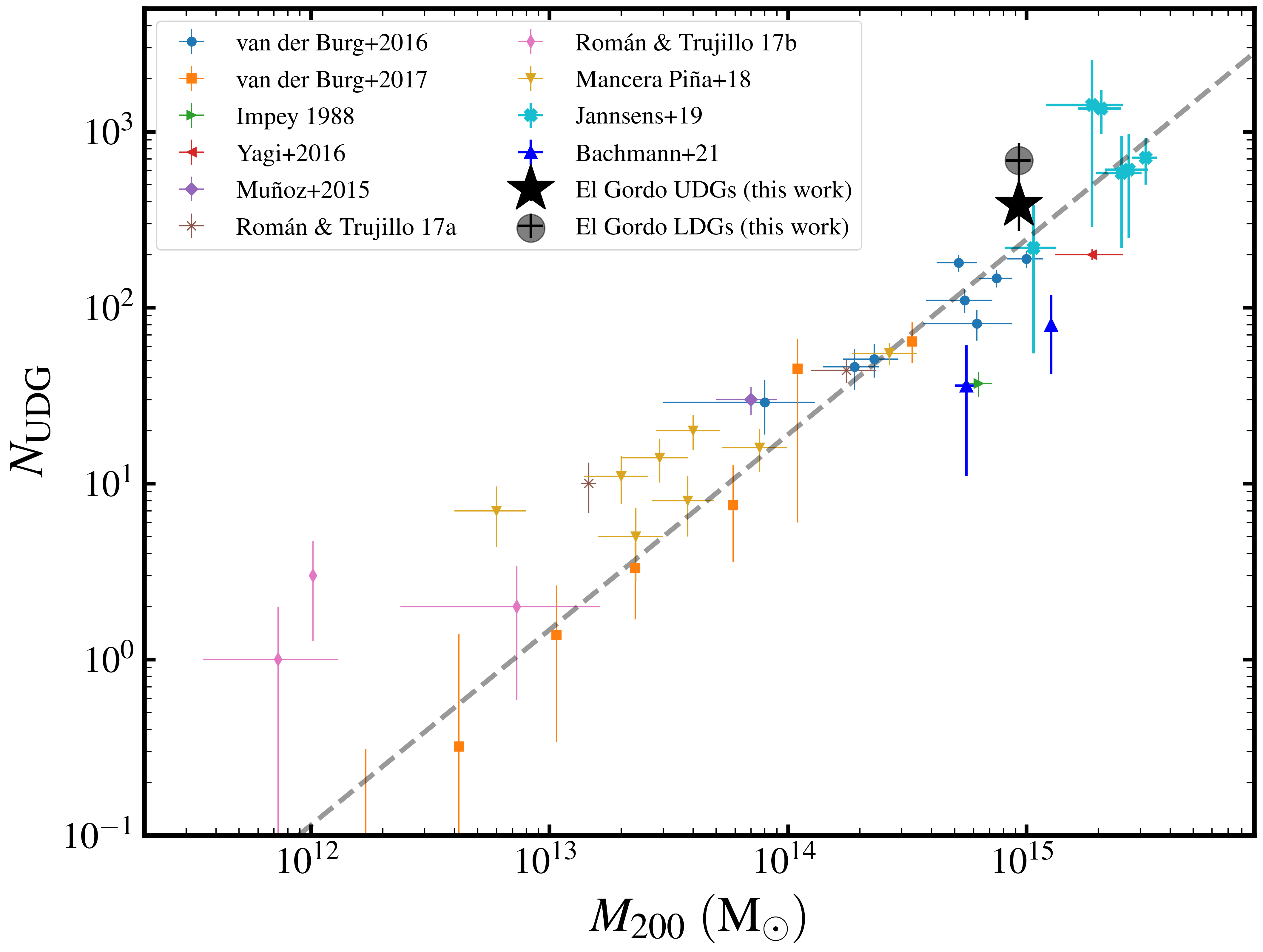}
	\caption{The abundance of surface-brightness selected UDGs in El Gordo from this work (black star) compared to other groups and clusters at lower redshift, with values from \citet[][blue points]{vanderBurg2016}, \citet[][orange squares]{vanderBurg2017}, \citet[][green right-pointing triangle]{Impey1988}, \citet[][red left pointing triangle]{Yagi2016}, \citet[][purple thick diamond]{Munoz2015}, \citet[][brown ``x"]{Roman2017a}, \citet[][pink diamond]{Roman2017b}, \citet[yellow downward-pointing triangles][]{ManceraPina2018}. The dashed line is the low-$z$ abundance-mass relation from \cite{vanderBurg2016}. We also show results from analysis of UDGs in the Hubble Frontier Fields at $z=0.3-0.55$ \citep[][cyan thick ``X"s]{Janssens2019} and clusters SPT-CL J2106-5844 ($z=1.13$) and MOO J1014-5844SPT-CL ($z=1.13$) from \citet[][blue upward pointing triangles]{Bachmann2021}. While selection effects complicate measurements of UDG abundance at high redshift, our measurement suggests that higher redshift clusters have a similar abundance as local clusters. Also shown is the abundance of stellar-surface-density selected LDGs as the large grey circle. The high abundance of those objects suggests that less than half of LDGs become UDGs at $z=0$.}
	\label{fig:nvsm}
\end{figure}

\subsection{Morphology}
\label{sec:morphology}
\begin{figure}
	\includegraphics[width=1\linewidth]{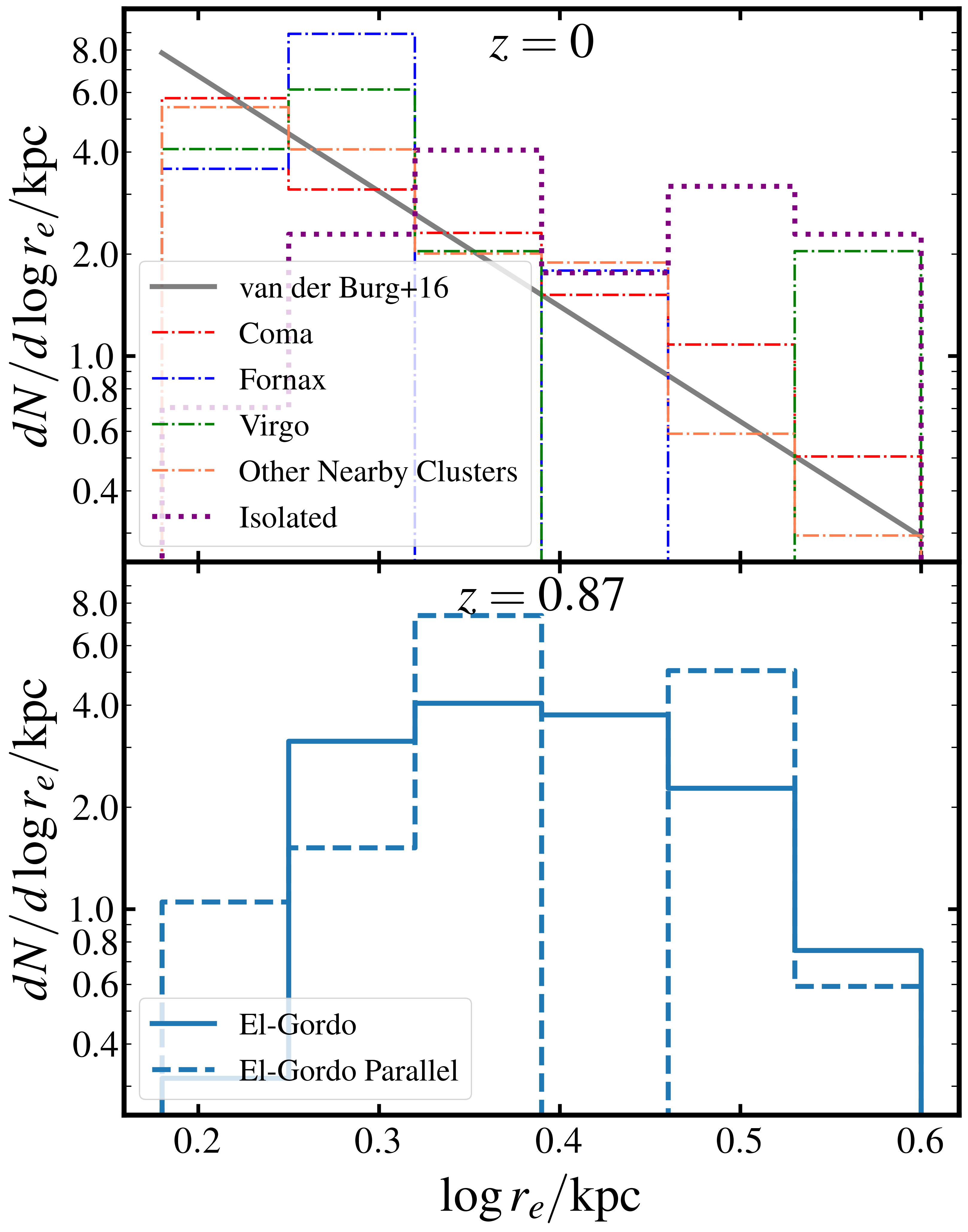}
	\caption{{\bf Top:} Size distribution of local cluster and isolated UDGs. The distribution of cluster UDGs is well characterized by a steeply falling ($n\propto r^{-3.4}$; \cite{vanderBurg2016}) size distribution (represented by the grey line), whereas isolated UDGs have a wider range of sizes. {\bf Bottom:} The size distribution of LDGs in the El Gordo cluster and parallel fields (solid and dashed lines respectively). Notably, the distribution of El Gordo LDGs have a flat size distribution (in both the cluster and parallel fields) that is more resemblant of local isolated UDGs than cluster objects.}
	\label{fig:sizedist}
\end{figure}

\begin{figure}
	\includegraphics[width=1\linewidth]{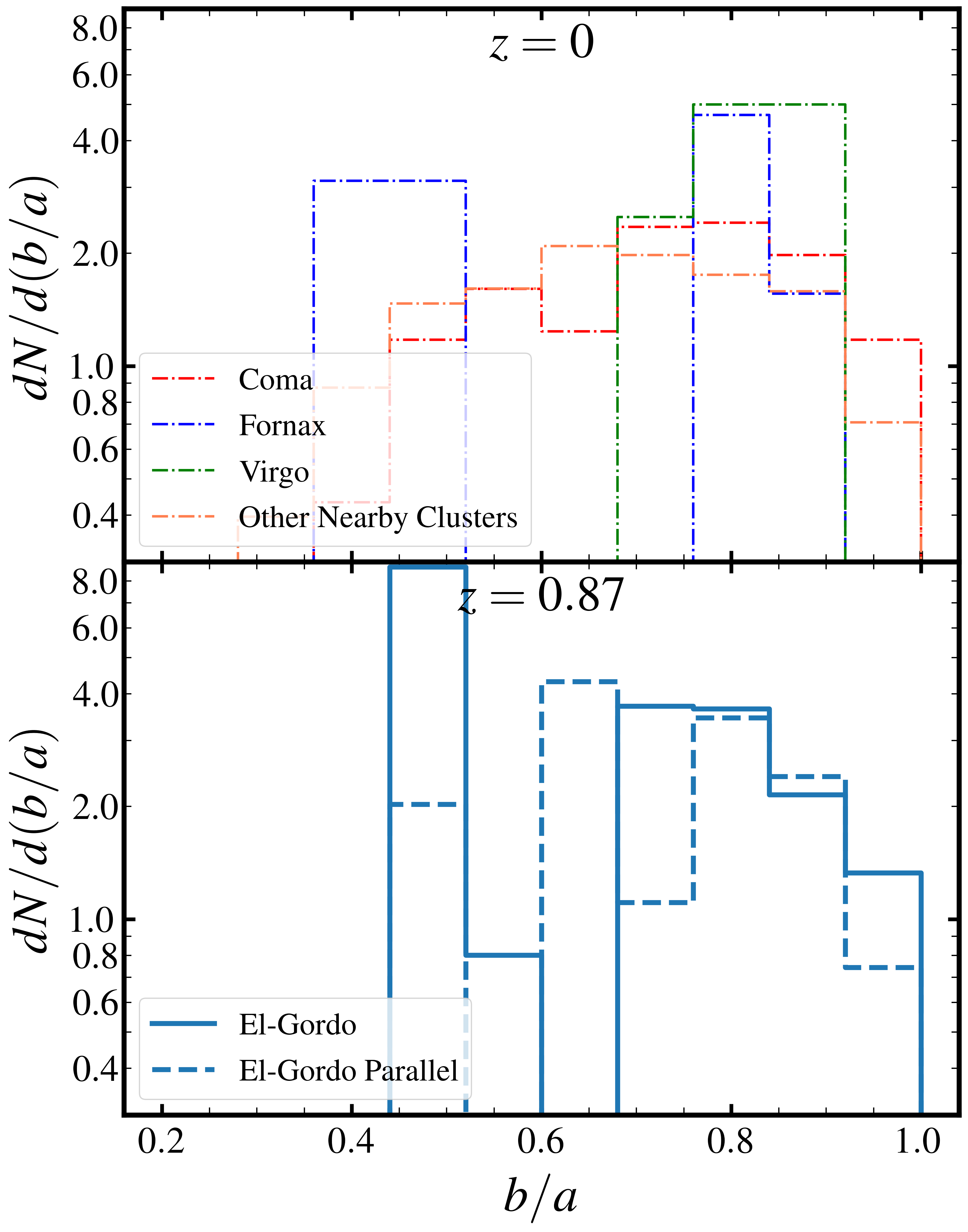}
	\caption{{\bf Top:} Axis-ratio distribution of local UDGs. {\bf Bottom:} The same for objects in El Gordo. UDGs identified in low-$z$ clusters have a relatively flat axis ratio distribution consistent with oblate spheroids, and LDGs in the El Gordo cluster field largely match this distribution, suggesting that they are consistent with being the progenitors of local UDGs and not much evolution of axis ratios happens between $z=0.87$ and today. }
	\label{fig:ardist}
\end{figure}

\begin{figure}
	\includegraphics[width=1\linewidth]{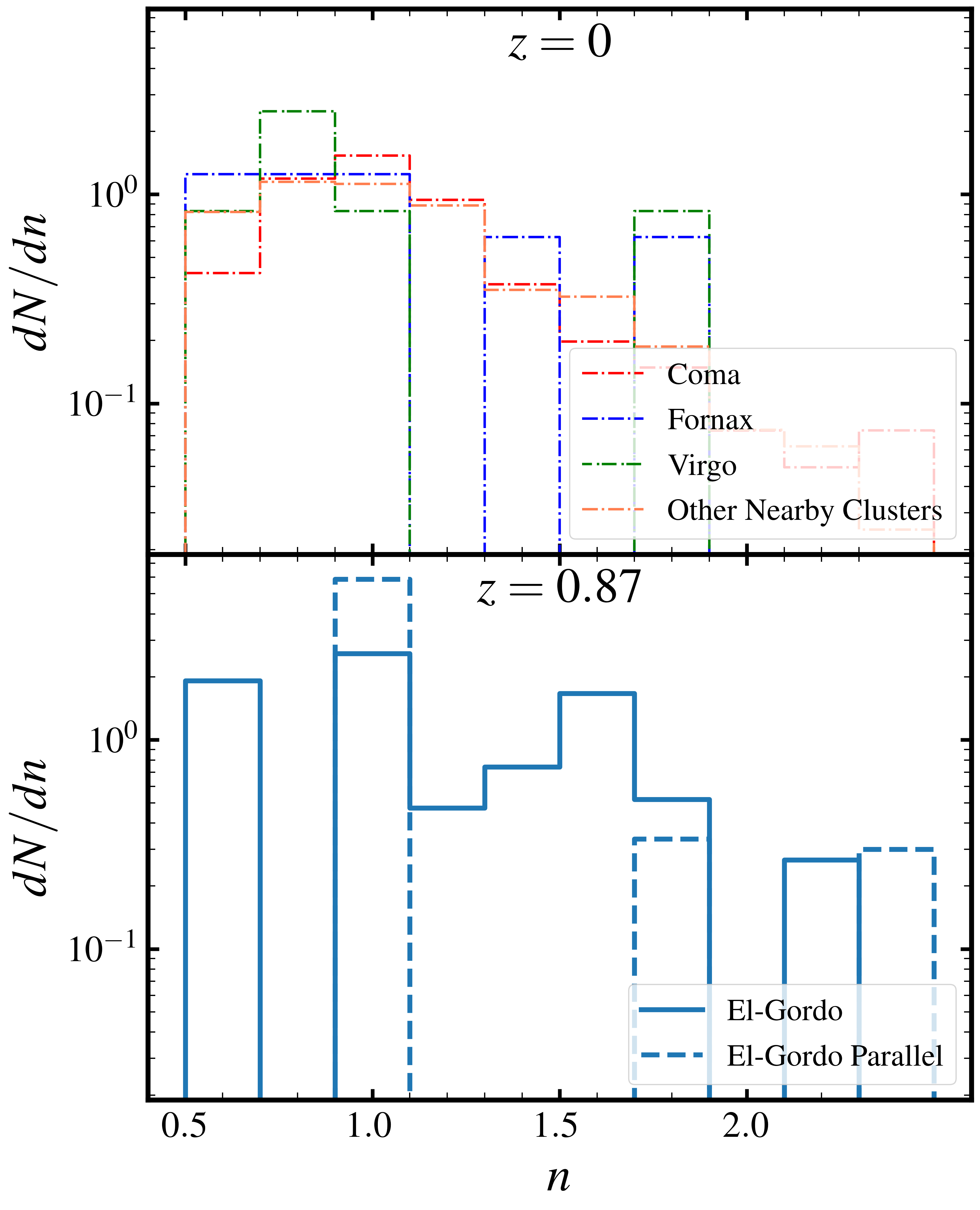}
	\caption{{\bf Top:} The S{\'e}rsic index distribution of low-$z$ UDGs. {\bf Bottom:} The same for El Gordo LDGs. The S{\'e}rsic indices of El Gordo LDGs are mainly $0.5-1.5$, consistent with local objects, suggesting that little evolution in $n$ is required between $z=0.87$ and $z=0$.}
	\label{fig:ndist}
\end{figure}
While local cluster UDGs have a range of sizes and axis ratios, they are dominated by small, round objects \citep{Koda2015,vanderBurg2016}. Among local cluster UDGs, $83\%$ have sizes less than $2.5$ kpc, following a steep $n\propto r^{-3.4}$ size distribution \citep{vanderBurg2016}. On the contrary, as shown in Fig.~\ref{fig:sizedist}\footnote{In Fig.~\ref{fig:sizedist}---\ref{fig:colordist}, the distributions of El Gordo and parallel field are weighted by the detection ability of objects (from Fig. ~\ref{fig:completeness}) and subtracted by the weighted, area-scaled, and lensing corrected distributions from the TDF.}, LDGs in El Gordo have a much flatter size distribution, consistent with $n\propto r^{0.8}$. This more so resembles the size distribution of the field UDG population (from \citealt{Leisman2017}) --- which rises between $1.5$~kpc and $3$~kpc --- before falling steeply after $3$ kpc, than local UDGs in Coma \citep{Yagi2016}, Fornax \citep{Venhola2017}, or Virgo \citep{Ferrarese2020}.

The axis ratio distribution of LDG in El Gordo, shown in Fig.~\ref{fig:ardist}, is more similar to local cluster objects. Local UDGs have an axis ratio distribution that peaks for larger $b/a$ values, consistent with oblate, stretched objects \citep[][but see \citealt{Rong2020}]{Burkert2017}. El Gordo LDGs are also largely consistent with this axis ratio distribution, suggesting that LDGs are consistent with being progenitors of local UDGs.

Lastly, we note that the S{\'e}rsic indices of LDGs in our sample (shown in Fig.~\ref{fig:ndist}) have a median S{\'e}rsic index of $1.0$, with a $25-75$ percentile range of $0.6-1.5$. This further suggests that LDGs in our sample are analogs of low-$z$ UDGs and validates the assumptions in our completeness tests in Sec.~\ref{sec:detection}. Although this is partially an artifact of our selection of objects with $n<2.5$, the $\sim20$ objects in our stellar surface density and size range in the cluster field with $n>2.5$ are primarily composed of $n=4$ objects unlike local UDGs.

\subsection{Colors and Stellar Populations}
\label{sec:colors}
\begin{figure}
	\includegraphics[width=1\linewidth]{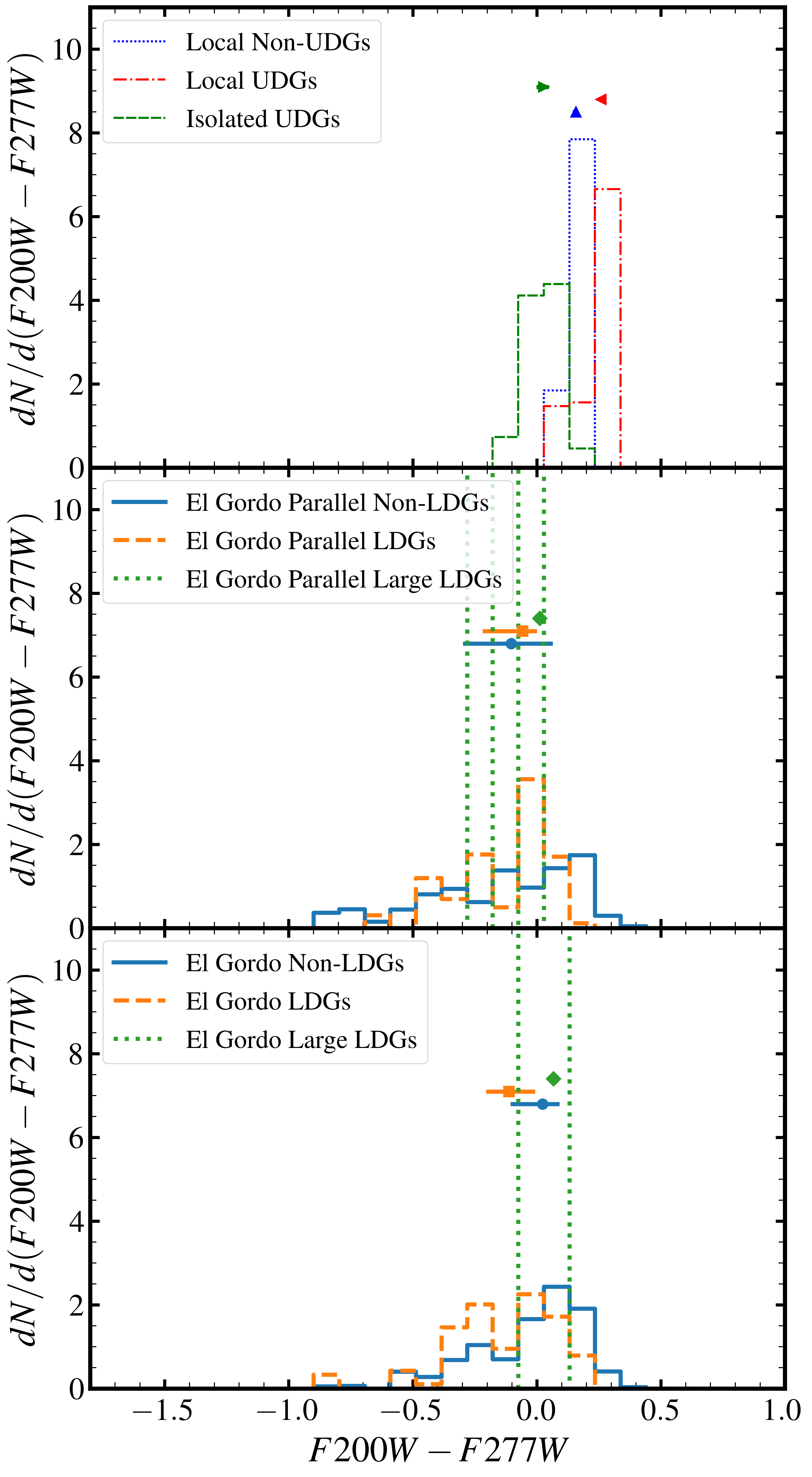}
	\caption{{\bf Top:} The color distribution of local cluster UDGs (red dot-dashed line), cluster non-UDGs (blue dotted line), and isolated UDGs (green dashed line). The left-facing, upward, and right-facing triangles and error bars (where larger than the points) correspond to the median and 32-68 percentile range of the respective distributions. Measured $g-i$ colors from Fornax, $B-R$ colors from Coma, and $g-z$ colors from Virgo have been translated to $F200W-F277W$ colors at $0.87$ assuming a single stellar population with a metallicity of $-0.79$ (see Sec.~\ref{sec:detection}). {\bf Middle:} The observed $F200W-F277W$ color distribution for non-LDGs (blue solid), LDGs (orange dashed) and large LDGs (green dotted, defined to be larger than $0.4\arcsec{}=4.8$~kpc). The blue circle, orange square, and green diamond with error bars correspond to the median and 32-68 percentile ranges of the distributions. {\bf Bottom:} Same as the middle panel, but for the El Gordo cluster field. While most cluster objects are red, consistent with old stellar populations, LDSGs and large LDGs tend to be \emph{bluer} than their non-LDG counterparts. This is the opposite of what is seen in local clusters, where UDGs are redder than non-UDGs. Instead, LDGs tend to more-so resemble isolated UDGs. Furthermore, the colors of LDGs in the cluster field are more similar to LDGs in the parallel field, and isolated UDGs at $z=0$, than other cluster dwarf galaxies.}
	\label{fig:colordist}
\end{figure}

In this section, we discuss the colors of LDGs in El Gordo. 
Figure~\ref{fig:colordist} show the distribution of F200W-F277W (rest-frame $Y-H$) color distribution of LDGs, non-LDGs, and large LDGs. The corresponding points and error-bars show the median and $32-68$ percentile ranges of the distributions. This is also compared with the distribution of the colors of local cluster UDGs. A population of red galaxies is apparent in all distributions, confirming that both LDGs and non-LDGs follow a red sequence associated with the cluster. However, LDG have a much more substantial population of blue galaxies.\footnote{We use this color because it is the reddest available color generally sensitive to the age of a stellar population. Bluer colors generally have more scatter in their distributions, but none show cluster LDGs as significantly redder than non-LDGs and the colors of cluster LDGs are usually similar to the colors of LDGs in the parallel field.}

Assuming that this color difference results from a stellar age difference, these results are consistent with results from Section~\ref{sec:morphology} suggesting that LDGs in El Gordo more resemble field objects than old, metal-poor objects observed in local clusters. (note that the expected evolution in color from a single metal-poor population from $3$~Gyr to $10$~Gyr is $\sim0.05$~mag, so passive evolution alone can't explain why LDGs in El-Gordo have relatively blue colors). This is further supported by the color distribution of LDGs in the parallel field, which is remarkably similar to the color distribution of LDGs in the cluster field, and by the best-fit spectral energy distributions, which suggest that LDGs have younger ages than non-LDGs. 

\vspace{5em}
\section{Discussion}
\label{sec:discussion}
Despite selecting objects that are expected to evolve into local UDGs, we find notable differences between LDGs in El Gordo and those in local clusters. Specifically, LDGs in El Gordo  tend to have a flatter size distribution than local UDGs, and as opposed to local UDGs, tend to be bluer than non-LDGs. When studying their sizes or colors, LDGs in the El Gordo cluster field are more similar to those in the parallel field than local cluster UDGs. Furthermore, there is not an excess or deficit of LDG in the cluster center compared to non-LDGs, rather the spatial distribution of the LDG population largely mirrors the overall galaxy population, and we find that the abundance of LDGs in El Gordo is slightly elevated compared to local clusters of similar mass. 

Altogether, this evidence suggests that LDGs in El Gordo fell into the cluster relatively recently --- they are largely similar to large isolated galaxies that have recently been accreted by the cluster. While many cluster UDGs at $z=0$ are old \citep{Gu2018,Ferre-Mateu2018,Ruiz-Lara2018,Pandya2018,Buzzo2022,Webb2022} some are young \citep{Rong2020,Buzzo2022}. On the other hand, our analysis seems to suggest that most LDGs in El Gordo are young. The quick LDG formation in El Gordo is consistent with processes like ram-pressure stripping \citep{Safarzadeh2017} or passive evolution of extended isolated galaxies \citep{Chan2018}. Furthermore, the redshift evolution in UDG properties from predominantly young with a flat size distribution to predominantly old with a steep size distribution suggests that environmental effects such as tidal heating \citep{Carleton2019} become more relevant by $z=0$.

\section{Conclusions}
\label{sec:conclusion}
We have identified the population of low surface density galaxies (LDGs), which are consistent with being the progenitors of $z=0$ UDGs assuming passive evolution, in the El Gordo cluster at $z=0.87$. We believe defining this sample of objects based on stellar mass surface density rather than surface brightness results in a more self-consistent selection criteria than observed surface brightness, which can vary significantly band-to-band. Based on our analysis of these objects, our conclusions are as follows.

\begin{itemize}
    \item While LDGs have similar morphologies (in terms of their axis ratios and S{\'e}rsic indices) as local UDGs, there are some notable differences between the high-$z$ LDG population and $z=0$ UDGs. In particular, the size distribution and colors of high-$z$ LDGs are more consistent with isolated UDGs than cluster UDGs at $z=0$, suggesting that LDGs in El Gordo experienced relatively little evolution since they were accreted and that further evolution is required to transform LDGs at $z=0.87$ to UDGs at $z=0$.
    \item The abundance of LDGs is slightly elevated compared with local UDGs, suggesting that some LDGs will indeed passively evolve into $z=0$ UDGs, but many will be tidally destroyed between $z=0.87$ and $z=0$. 
    \item The spatial distribution of LDGs is consistent with other dwarfs and doesn't show a deficit in the cluster center. This further suggests that tidal destruction of LDGs is significant between $z=0.87$ and $z=0$.
\end{itemize}

\vspace{1em}
\noindent{\bf Acknowledgements:}
%
We are grateful to the anonymous referee for their helpful comments. This work is based on observations made with the NASA/ESA/CSA James Webb Space
Telescope. The data were obtained from the Mikulski Archive for Space Telescopes
(MAST) at the Space Telescope Science Institute, which is operated by the
Association of Universities for Research in Astronomy, Inc., under NASA contract
NAS 5-03127 for {\it JWST}. These observations are associated with {\it JWST} programs 1176 and 2738.

We dedicate this paper to the memory of our dear PEARLS colleague Mario Nonino, who was a gifted and dedicated scientist, and a generous person.
TMC is grateful for support from the Beus Center for Cosmic Foundations. RAW, SHC, and RAJ acknowledge support from NASA {\it JWST} Interdisciplinary
Scientist grants NAG5-12460, NNX14AN10G and 80NSSC18K0200 from GSFC. J.M.D. acknowledges the support of project PGC2018-101814-B-100 (MCIU/AEI/MINECO/FEDER, UE) Ministerio de Ciencia, Investigaci\'on y Universidades.  This project was funded by the Agencia Estatal de Investigaci\'on, Unidad de Excelencia Mar\'ia de Maeztu, ref. MDM-2017-0765. CC is supported by the National Natural Science Foundation of China, No. 11803044, 11933003, 12173045. This work is sponsored (in part) by the Chinese Academy of Sciences (CAS), through a grant to the CAS South America Center for Astronomy (CASSACA). We acknowledge the science research grants from the China Manned Space Project with NO. CMS-CSST-2021-A05. RAB gratefully acknowledges support from the European Space Agency (ESA) Research Fellowship. CJC acknowledges support from the European Research Council (ERC) Advanced Investigator
Grant EPOCHS (788113). CNAW acknowledges funding from the {\it JWST}/NIRCam contract NASS-0215 to the University of Arizona. MAM acknowledges the support of a National Research Council of Canada Plaskett Fellowship, and the Australian Research Council Centre of Excellence for All Sky Astrophysics in 3 Dimensions (ASTRO 3D), through project number CE17010001.

We also acknowledge the indigenous peoples of Arizona, including the Akimel
O'odham (Pima) and Pee Posh (Maricopa) Indian Communities, whose care and
keeping of the land has enabled us to be at ASU's Tempe campus in the Salt
River Valley, where much of our work was conducted. 

\software{ Astropy: \url{http://www.astropy.org} \citep{Robitaille2013,
Astropy2018}; Photutils:
\url{https://photutils.readthedocs.io/en/stable/} \citep{Bradley2020};
Profound: \url{https://github.com/asgr/ProFound} \citep{Robotham2017, 
Robotham2018}; ProFit: \url{https://github.com/ICRAR/ProFit}
\citep{Robotham2018}; {\sc GALFIT}: \url{https://users.obs.carnegiescience.edu/peng/work/galfit/galfit.html} \citep{Peng2010}; SourceExtractor:
\url{https://sextractor.readthedocs.io/en/latest/} \citep{Bertin1996}; Python FSPS: \url{https://dfm.io/python-fsps/current/} \citep{Conroy2009,Conroy2010,Foreman-Mackey2014}; Prospector: \url{https://prospect.readthedocs.io/en/latest/} \citep{Johnson2021}; WebbPSF: \url{https://webbpsf.readthedocs.io/en/latest/}.}
\facilities{James Webb Space Telescope; Mikulski Archive
\url{https://archive.stsci.edu}.}\

\bibliography{eg_udg_references}

\end{document}